\documentclass[journal, 12pt, onecolumn, draftclsnofoot]{IEEEtran}
\usepackage[utf8]{inputenc} 
\usepackage[T1]{fontenc}
\usepackage{url}
\usepackage{ifthen}
\usepackage{cite}
\usepackage[cmex10]{amsmath} 
\usepackage[hidelinks]{hyperref}
\usepackage{tikz}
\usetikzlibrary{calc}
\usepackage{multicol}
\usepackage{mdframed}
\usepackage{color}
\usepackage{epsfig,subfigure}
\usepackage{amssymb, amsmath}
\usepackage{color}
\usepackage{bm}
\usepackage{amsthm}
\usepackage{authblk}
\usepackage{cite}
\usetikzlibrary{shapes,decorations,arrows,calc,arrows.meta,fit,positioning}
\newcommand{\define}{\stackrel{\triangle}{=}}

\usepackage{graphicx}
\usepackage{color}

\usepackage{amssymb}
\usepackage{amsmath,amsfonts,amssymb}
\usepackage{verbatim}
\usepackage{stfloats}
\usepackage[bookmarks=false]{}

\DeclareMathOperator{\pr}{Pr}

\newtheorem{theorem}{Theorem}
\newtheorem{corollary}{Corollary}
\newtheorem{definition}{Definition}
\newtheorem{lemma}{Lemma}

\newtheorem{remark}{Remark}

\newcommand{\indep}{\perp \!\!\! \perp}

\begin{document}

\title{\Large Double Blind $T$-Private Information Retrieval}
\author{\normalsize Yuxiang Lu, Zhuqing Jia and Syed A. Jafar}
\affil{\small Center for Pervasive Communications and Computing (CPCC), UC Irvine\\
Email: \{yuxiang.lu, zhuqingj, syed\}@uci.edu}
\date{}
\maketitle

\begin{abstract}
Double blind $T$-private information retrieval (DB-TPIR) enables two users, each of whom specifies an index ($\theta_1, \theta_2$, resp.), to efficiently retrieve a message $W(\theta_1,\theta_2)$ labeled by the two indices, from a set of $N$ servers that store all messages $W(k_1,k_2), k_1\in\{1,2,\cdots,K_1\}, k_2\in\{1,2,\cdots,K_2\}$, such that the two users' indices are kept private from any set of up to $T_1,T_2$ colluding servers, respectively, as well as from each other. A DB-TPIR scheme based on cross-subspace alignment is proposed in this paper, and shown to be capacity-achieving in the asymptotic setting of large number of messages and bounded latency. The scheme is then extended to $M$-way blind $X$-secure $T$-private information retrieval (MB-XS-TPIR) with multiple  ($M$) indices, each belonging to a different user, arbitrary privacy levels for each index ($T_1, T_2,\cdots, T_M$), and arbitrary level of security ($X$) of data storage, so that the message $W(\theta_1,\theta_2,\cdots, \theta_M)$ can be efficiently retrieved while the stored data is held secure against collusion among up to $X$ colluding servers, the $m^{th}$ user's index is private against collusion among up to $T_m$ servers, and each user's index $\theta_m$ is private from all other users. The general scheme relies on a tensor-product based extension of cross-subspace alignment and retrieves $1-(X+T_1+\cdots+T_M)/N$ bits of desired message per bit of download.
\end{abstract}

\section{Introduction}
Data privacy and security are among the biggest challenges of the modern information age. Driven by these challenges there is much interest in the  building blocks (primitives) of privacy/security preserving schemes, such as secret sharing \cite{Shamir}, oblivious transfer \cite{Gertner_Goldwasser_Malkin}, private information retrieval (PIR) \cite{PIRfirst, PIRfirstjournal}, secure multiparty computation (MPC) \cite{yao1982protocols,Yao_2PC,Goldreich_Micali_Wigderson_MPC}, and private simultaneous messages (PSM) \cite{Feige_Killian_Naor_PSM}. Understanding the fundamental limits of each of these building blocks is the key to understanding the scope of their potential applications. The focus of this work is on private information retrieval (PIR).

Introduced by Chor et al.  in \cite{PIRfirst, PIRfirstjournal}, the goal of PIR in its simplest form is to allow a user to  efficiently retrieve a desired message from a set of $K$ messages that are replicated across $N$ distributed servers, while revealing no information to any individual server about which message is desired.  Until recently, PIR was investigated primarily by computer scientists and cryptographers \cite{PIRfirst, PIRfirstjournal} under the assumption of \emph{short} messages (e.g.,  each message is  just one bit), with the goal of minimizing the total communication (upload and download) cost. However, following the capacity characterization of PIR in \cite{Chan_Ho_Yamamoto, Sun_Jafar_PIR} under the assumption of \emph{long} messages (where downloads dominate the communication cost), the fundamental limits (capacity) of various forms of download-efficient PIR have become an active topic in information theory. Recent advances include the capacity characterizations of PIR with $T$-privacy \cite{Sun_Jafar_TPIR}, symmetric-privacy \cite{Sun_Jafar_SPIR}, weak privacy\cite{Samy_Attia_Tandon_Lazos_Weak,Lin_Kumar_Rosnes_Eitan_Weak}, eavesdroppers and/or Byzantine servers \cite{Wang_Skoglund_SPIREve,Wang_Sun_Skoglund, Wang_Skoglund_PIRSPIRAd,Wang_Sun_Skoglund_BSPIR, Banawan_Ulukus_BPIR}, coded storage \cite{FREIJ_HOLLANTI,Tajeddine_Gnilke_Karpuk, Wang_Skoglund_TSPIR,Sun_Jafar_MDSTPIR, Tajeddine_Gnilke_Karpuk_Hollanti, Wang_Skoglund_MDS, Jia_Jafar_MDSXSTPIR,Zhou_Tian_Sun_Liu_Min_Size}, secure storage \cite{Yang_Shin_Lee,Jia_Sun_Jafar_XSTPIR,Jia_Jafar_GXSTPIR}, limited storage \cite{Attia_Kumar_Tandon,Wei_Arasli_Banawan_Ulukus_Decentralized,Woolsey_Chen_Ji_Storage,Guo_Zhou_Tian_Storage,Banawab_Arasli_Wei_Ulukus_Heterogeneous}, cached data or side information \cite{Tandon_CachePIR, Wei_Banawan_Ulukus_Side,Wei_Banawan_Ulukus, Chen_Wang_Jafar_Side}, multiple rounds \cite{Sun_Jafar_MPIR,Yao_Liu_Kang_Multiround}, multiple desired messages \cite{Banawan_Ulukus_MPIR, Shariatpanahi_Siavoshani_Maddah, Jia_Jafar_SDMM, Wang_Banawan_Ulukus_PSI}, upload constraints \cite{Tian_Sun_Chen_Upload}, arbitrary collusion patterns \cite{Tajeddine_Gnilke_Karpuk,Yao_Liu_Kang_Collusion_Pattern}, single server PIR with user side information \cite{Li_Gastpar, Li_Gastpar_SSMUPIR,Alex_Single_1,Alex_Single_2,Alex_Single_3,Alex_Single_4,Alex_Single_5}, latent-variable single server PIR\cite{Latent_PIR}, as well as applications of PIR to private computation \cite{Mirmohseni_Maddah,Sun_Jafar_PC,Mousavi_MaddahAli_Mirmohseni_Innerproduct, Obead_Lin_Rosnes_Kliewer_PFC}, private search \cite{Chen_Wang_Jafar_Search}, private set intersection \cite{Wang_Banawan_Ulukus_PSI}, coded computing \cite{Jia_Jafar_CDBC},  locally decodable codes \cite{Sun_Jafar_LDC}, etc.

Our goal in this work is to further expand the understanding of download-efficient PIR in a new direction --- \emph{$M$-way blind} $X$-secure $T$-PIR or MB-XS-TPIR, where the data, labeled by $M$ indices, is  stored in an $X$-secure\footnote{$X$-security ($T$-privacy) means that  security (privacy) is guaranteed against any set of up to $X$ ($T$) colluding servers.} fashion by $N$ servers, and $M$ users jointly retrieve a desired message by specifying one index each (user $m$ specifies $\theta_m$, $\forall m\in\{1,2,\cdots, M\}$), while keeping their  index private from each other and also $T$-private from the servers where the data is stored. It is conceivable that such a functionality may be directly useful. For example, consider private data, e.g., health records, that are stored anonymously and $X$-securely among a cloud of distributed servers. For enhanced security it is not uncommon to require multi-factor authentication, e.g., 2-factor authentication from a pair of devices (say, smartphone and computer) that belong to the owner of the data (patient) in order to allow access to the data. This can be  implemented as the double blind setting of MB-XS-TPIR by creating $2$ passwords (indices $\theta_1, \theta_2$), so that the two devices must each provide $\theta_1, \theta_2$ respectively, in order for the patient  to retrieve $\mathbf{W}(\theta_1, \theta_2)$ on either device. It is important that each device learns nothing about the other device's password (treating devices as users, this is called inter-user privacy), so that the loss or hacking of either device does not reveal more than its own password. Furthermore, the passwords/indices are also kept $T$-private from the servers, so that even the servers learn nothing about which record is being retrieved. $M$-way authentication similarly motivates   MB-XS-TPIR. In general, MB-XS-TPIR may be a good solution for secret sharing among multiple parties when the size of the secret is too large so that it needs to be securely stored among distributed servers (cloud) while access to the secret is allowed by distributing smaller  keys or passwords (indices in MB-XS-TPIR) to the parties. The multiway blind functionality  is also useful for secure multiparty computation\footnote{A notable limitation is that $M$-way blind PIR allows  communication only between users and servers, but Secure MPC protocols may in general also allow direct communication between users.} where the inputs $\theta_1, \cdots, \theta_M$ of a function $f(x_1, \cdots, x_M)$ are held by $M$ parties and $\mathbf{W}$, whose $(\theta_1, \cdots, \theta_M)^{th}$ entry is the evaluation of the function at $(\theta_1, \cdots, \theta_M)$, is stored by distributed servers \cite{Ishai_Kushilevitz}. Fundamentally, however, our motivation is simply to expand the scope of a basic primitive.

The main contribution of this work is a cross-subspace alignment (CSA) based scheme for MB-XS-TPIR. To place this in perspective, we note that the evolution of CSA codes has followed a remarkable trajectory with  crossovers between PIR and coded distributed computing (CDC). In a nutshell, CSA codes originated in PIR, then crossed over to CDC where the  constructions were generalized, and now in this work, return back to PIR in their generalized form which allows MB-XS-TPIR. To see this in a bit more detail, recall that the idea of cross-subspace alignment originated in the context of XS-TPIR \cite{Jia_Sun_Jafar_XSTPIR, Jia_Jafar_GXSTPIR} as a way to align interference from undesired product terms that result when a secret-shared (private) query vector is multiplied with a secret-shared (secure) data vector. It was then observed in  \cite{Kakar_Ebadifar_Sezgin_CSA, Jia_Jafar_SDMM, Jia_Sun_Jafar_XSTPIR,Jia_Jafar_CDBC, Chen_Jia_Wang_Jafar_NGCSA}  that the idea of aligning undesired product terms is similarly useful in  distributed computing applications, which led to  a crossover of CSA codes to coded distributed computing \cite{Cadambe_Grover_Tutorial}. Generalized CSA codes were constructed in \cite{Jia_Jafar_CDBC} to unify and improve upon several state-of-art CDC approaches like Lagrange Coded Computing \cite{Yu_Lagrange} and Entangled Polynomial codes \cite{Yu_Maddah-Ali_Avestimehr_Polynomial}. The generalized forms of CSA codes allow not only  pairwise matrix multiplications, but also  multilinear computations. This work represents the next step forward, as the generalizations of CSA codes that emerged in the context of coded distributed computing are used to enable new forms of PIR. Indeed, the main idea behind this work is the framing of a  particular solution\footnote{ The problem of MB-XS-TPIR, or PIR in general,  is not  \emph{equivalent} to distributed matrix (tensor) multiplication. For example, there is no constraint in PIR that forces the answers returned by the servers to be linear in either the query vectors or the stored information, or more specifically, products of query vectors and the stored information. However, many \emph{solutions} to PIR indeed take this form, thus creating a connection between PIR and CDC. That such solutions tend to be optimal in many cases strengthens this connection.} to MB-XS-TPIR as a problem of distributed secure tensor product computation. With this mapping we find that the key to the solution is to compute the tensor products of suitably structured secret-shared query vectors that originate at the users, and correspondingly structured secret-shared data matrices that are stored at the servers.  Note that CSA codes allow a range of structures corresponding to various choices of feasible code parameters, which may be further optimized for download cost depending on the application. See Section \ref{sec:genscheme} for additional details.  The desired tensor-products turn out to be multilinear operations, so that the multilinear computation capability of CSA codes can be applied to  MB-XS-TPIR.

In order to introduce our solution in a more transparent setting, our initial focus  is on DB-TPIR, i.e., the double-blind setting $(M=2)$ with $T$-private user indices ($T_1,T_2$, resp.) and replicated data storage, initially with no data-security, i.e., $X=0$. This basic setting allows us to  convey the main ideas behind the construction of the scheme and also to explore its optimality. Specifically, for the DB-TPIR problem we propose a scheme based on cross-subspace alignment \cite{Jia_Jafar_CDBC} which allows the retrieval of $1-(T_{1}+T_{2})/N$ bits of desired message per bit of download, regardless of the number of messages. By noting connections between this problem and $X$-secure $T$-private information retrieval (XS-TPIR) \cite{Jia_Sun_Jafar_XSTPIR} we show that $1-(T_{1}+T_{2})/N$ is also the  asymptotic capacity of DB-TPIR as the number of messages approaches infinity, provided that the number of bits of each message that are jointly encoded is bounded (say, due to latency constraints). 

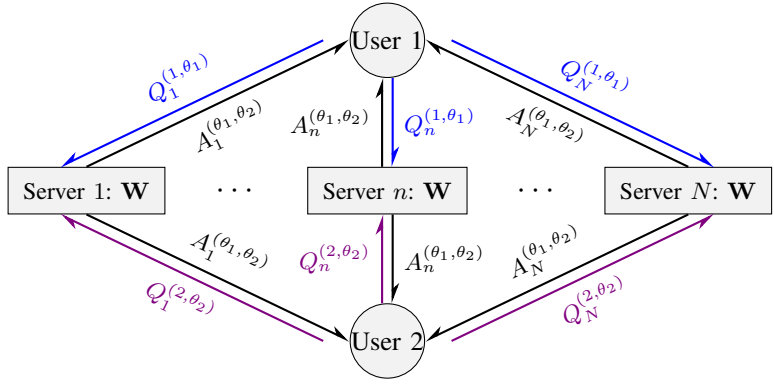
\begin{figure}[!h]
\centering
\begin{tikzpicture}
\node [draw, rectangle,fill=black!5, text=black, inner sep =0.2cm] (S1) at (4cm, 0cm) {\footnotesize Server $1$: $\mathbf{W}$};
\node [rectangle, inner sep =0.2cm] (Ddots1) at (6cm, 0cm) {$\cdots$};
\node [draw, rectangle, fill=black!5, text=black, inner sep =0.2cm] (Sn) at (8cm, 0cm) {\footnotesize Server $n$: $\mathbf{W}$};
\node [rectangle, inner sep =0.2cm] (Ddots2) at (10cm, 0cm) {$\cdots$};
\node [draw, rectangle, fill=black!5, text=black, inner sep =0.2cm] (SN) at (12cm, 0cm) {\footnotesize Server $N$: $\mathbf{W}$};

\node[circle, draw=black, fill=black!5, text=black, minimum size=0.9cm, inner sep=0] (U1) at (8cm, 2cm) { \small User 1};
\node[circle, draw=black, fill=black!5, text=black, minimum size=0.9cm, inner sep=0] (U2) at (8cm, -2cm) { \small User 2};

\draw [blue, thick, arrows = {-Stealth[right]}] ([xshift=-10pt]U1.west) to node[above, sloped] {\footnotesize $Q_{1}^{(1,\theta_{1})}$} ([xshift=-10pt]S1.north);
\draw [blue, thick, arrows = {-Stealth[left]}] ([xshift=2pt]U1.south) to node[right] {\footnotesize $Q_{n}^{(1,\theta_{1})}$} ([xshift=2pt]Sn.north);
\draw [blue, thick, arrows = {-Stealth[left]}] ([xshift=10pt]U1.east) to node[above, sloped] {\footnotesize $Q_{N}^{(1,\theta_{1})}$} ([xshift=10pt]SN.north);

\draw [violet, thick, arrows = {-Stealth[left]}] ([xshift=-10pt]U2.west) to node[below, sloped] {\footnotesize $Q_{1}^{(2,\theta_{2})}$} ([xshift=-10pt]S1.south);
\draw [violet, thick, arrows = {-Stealth[left]}] ([xshift=-2pt]U2.north) to node[left] {\footnotesize $Q_{n}^{(2,\theta_{2})}$} ([xshift=-2pt]Sn.south);
\draw [violet, thick, arrows = {-Stealth[right]}] ([xshift=10pt]U2.east) to node[below, sloped] {\footnotesize $Q_{N}^{(2,\theta_{2})}$} ([xshift=10pt]SN.south);

\draw [black, thick, arrows = {-Stealth[right]}] (S1.north) to node[below, sloped] {\footnotesize $A_{1}^{(\theta_{1}, \theta_{2})}$} (U1.west);
\draw [black, thick, arrows = {-Stealth[left]}] ([xshift=-2pt]Sn.north) to node[left] {\footnotesize $A_{n}^{(\theta_{1}, \theta_{2})}$} ([xshift=-2pt]U1.south);
\draw [black, thick, arrows = {-Stealth[left]}] (SN.north) to node[below, sloped] {\footnotesize $A_{N}^{(\theta_{1}, \theta_{2})}$} (U1.east);

\draw [black, thick, arrows = {-Stealth[left]}] (S1.south) to node[above, sloped] {\footnotesize $A_{1}^{(\theta_{1}, \theta_{2})}$} (U2.west);
\draw [black, thick, arrows = {-Stealth[left]}] ([xshift=2pt]Sn.south) to node[right] {\footnotesize $A_{n}^{(\theta_{1}, \theta_{2})}$} ([xshift=2pt]U2.north);
\draw [black, thick, arrows = {-Stealth[right]}] (SN.south) to node[above, sloped] {\footnotesize $A_{N}^{(\theta_{1}, \theta_{2})}$} (U2.east);
\end{tikzpicture}
\caption{\it The double blind $T$-private information retrieval (DB-TPIR) problem. }
\label{fig:JTPIR}
\end{figure}

With the insights obtained from DB-TPIR, we are then able to fully generalize our achievable scheme to MB-XS-TPIR, i.e., $M$-way blind $X$-secure  $T$-private information retrieval with multiple ($M$) indices, each specified privately by a different user, arbitrary privacy levels for each index ($T_1, T_2,\cdots, T_M$), and arbitrary level of security ($X$) of data storage, so that the message $W(\theta_1,\theta_2,\cdots, \theta_M)$ can be efficiently retrieved by the users while the stored data is held secure against collusion among up to $X$ colluding servers, the $m^{th}$ user's index is private against collusion among up to $T_m$ servers, and each user's index $\theta_m$ is private from all other users. The general setting is based on an $M$-way tensor-product extension of cross-subspace alignment codes, and retrieves $1-(X+T_1+\cdots+T_M)/N$ bits of desired message per bit of download. This generalizes the known asymptotically (large number of messages) optimal schemes for various special cases of MB-XS-TPIR including DB-TPIR $(M=2, X=0)$ and XS-TPIR $(M=1)$ \cite{Jia_Sun_Jafar_XSTPIR} (which automatically recovers asymptotically optimal schemes for TPIR ($X=0, M=1)$ \cite{Sun_Jafar_TPIR} and PIR ($X=0, M=1,T_1=1$) \cite{Sun_Jafar_PIR} as well). In fact, the achievable scheme for MB-XS-TPIR also satisfies symmetric-privacy, i.e., the users learn nothing  about the database or each others' indices, beyond the desired message. Therefore,  it also yields symmetrically private schemes as special cases. For example, the general MB-XS-TPIR scheme yields a  capacity achieving scheme for Symmetric XS-TPIR ($M=1$) \cite{Chen_Jia_Wang_Jafar_NGCSA}, STPIR ($M=1, X=0$, Symmetric Privacy) \cite{Wang_Skoglund_TSPIR} and SPIR ($M=1, X=0, T_1=1)$ as well. Based on all these observations, we conjecture that the general MB-XS-TPIR scheme is  also asymptotically optimal.

In order to compare the new scheme with state of art, a natural baseline is obtained from \cite{Ishai_Kushilevitz} where a secure multiparty computation (MPC) scheme is constructed based on symmetric-PIR (SPIR) as a building block. This construction can be naturally generalized to a DBPIR scheme. Intuitively,  this construction is based on a partitioning of $N$ servers into $\sqrt{N}$ groups of $\sqrt{N}$ servers each, such that within each sub-group the SPIR scheme is executed for one user, while across sub-groups the SPIR scheme is executed for the other user. However, even with the most efficient SPIR scheme as the building block, the rate of this construction for DBPIR is $\left(1-1/\sqrt{N}\right)^2$, which is strictly smaller than the rate $1-2/N$ achieved by our asymptotically optimal scheme. This is because cross-subspace alignment allows us to avoid the $2$-way partitioning of servers and is able to gain significant efficiency by jointly exploiting all servers. For example, with $N=4$ servers, the partitioning based approach achieves a rate of $\left(1-1/\sqrt{N}\right)^2=1/4$, while the new scheme  achieves a $100\%$ higher rate of $1-2/N= 1/2$ due to cross-subspace alignment.

This paper is organized as follows. Section \ref{sec: form} formalizes the general MB-XS-TPIR problem. Section \ref{sec: main} states the main results of this paper in the form of two theorems. Their proofs are presented in Section \ref{sec: DB-TPIR} and Section \ref{sec: MB-XS-TPIR}. Section \ref{sec: conclusion} concludes the paper.

\emph{Notation:} For any two integers $a,b$ such that $a \leq b$, let $[a : b]$ denote the set $\{a, a+1, \cdots, b\}$. Let $X_{[a:b]}$ denote the set $\{X_{a}, X_{a+1}, \cdots, X_{b}\}$. For any index set $\mathcal{I} = \{i_{1}, i_{2}, \cdots, i_{n}\}$, $X_{\mathcal{I}}$ denotes the set $\{X_{i_{1}}, X_{i_{2}}, \cdots, X_{i_{n}}\}$. For two vectors $\mathbf{A}$ and $\mathbf{B}$, $\mathbf{A} \indep \mathbf{B}$ denotes that they are linearly independent. The notation $\mathbf{A}^{\prime}$ denotes the transpose of $\mathbf{A}$, and $\mathbf{A}(i)$ denotes the $i^{th}$ entry of $\mathbf{A}$. For an $n$-dimensional tensor $\mathbf{C}$, the notation $\mathbf{C}(i_{1}, i_{2}, \cdots, i_{n})$ represents the entry at the corresponding position of $\mathbf{C}$. If $\mathbf{C}$ is a two-dimensional tensor, then it is a matrix and $\mathbf{C}(i_{1},i_{2})$ denotes the $(i_{1},i_{2})^{th}$ entry of matrix $\mathbf{C}$. The notation $(x)^+$ denotes $\max(x,0)$. If $A$ is a set of random variables, then by $H(A)$ we denote the joint entropy of those random variables. Mutual information between sets of random variables are similarly defined with the notation $I(A;B)$. The notation $\mathbf{e}_{K}(\theta)$ denotes the $\theta^{th}$ column of the $K\times K$ identity matrix.

\section{Problem Statement: MB-XS-TPIR}\label{sec: form}
Consider a database $\mathbf{W}$ comprised of $K=K_{1}K_2\cdots K_M$ messages, indexed as 
\begin{align}
\mathbf{W}=\bigg(\mathbf{W}(k_1,k_2,\cdots, k_M)\bigg)_{k_1\in[1:K_1],\cdots,k_M\in[1:K_M]}.
\end{align}

Each message consists of  a stream of  i.i.d. uniform bits. The \emph{stream} of symbols implies that the message lengths are unbounded (a standard assumption in information theory). However, we are interested primarily in \emph{bounded-latency} MB-XS-TPIR schemes, i.e., schemes that code over a bounded number of bits. For example, consider an encoder that accepts as input  $L$ symbols from $\mathbb{F}_q$ for each message, i.e., $L\log_2(q)$ bits of each message, and jointly encodes them. In order to jointly encode its inputs, the encoder must first wait to collect $L\log_2(q)$ bits of data for each message, thus introducing a coding delay, or latency. By bounded latency,  we mean that $L, q$ are $O(1)$ in the parameters $K_1, K_2,\cdots, K_M$. In other words, the number of bits that are jointly encoded by the MB-XS-TPIR scheme is bounded even as the number of messages approaches infinity. This assumption  is important in practice, especially for streaming or dynamic data.  To our knowledge, for all PIR settings where the asymptotic (large number of messages) capacity is known, it is achieved by bounded-latency schemes \cite{Jia_Jafar_MDSXSTPIR}. So we do not expect the bounded latency assumption to affect the asymptotic capacity of MB-XS-TPIR. But it will be a useful assumption for converse arguments for the special case of DB-TPIR (Double Blind $T$-PIR). Another issue worth clarifying is that even though $L$ is bounded while the number of messages is allowed to be much larger, the downloads still dominate the communication cost because the same queries can be re-used repeatedly to download the unbounded desired message stream, $L$ symbols at a time.

Under the bounded latency assumption, without loss of generality we will assume that each message has length $L$ symbols. In $q$-ary units,
\begin{align}
H(\mathbf{W}(k_1,k_2,\cdots k_M)) = L,&& \forall k_1\in[1:K_1],\cdots,k_M\in[1:K_M],
\end{align}
\begin{align}
H(\mathbf{W}) = \sum_{k_1 \in [1:K_{1}], \cdots, k_M \in [1:K_{M}]}H\bigg(\mathbf{W}(k_1,k_2,\cdots k_M)\bigg) = K_{1}K_{2}\cdots K_{M}L.
\end{align}

The database $\mathbf{W}$ is  stored at $N$ distributed servers according to an $X$-secure storage scheme. Let the storage at the $n^{th}$ server be denoted by $\mathbf{S}_{n}, n\in[1:N]$. An $X$-secure storage scheme ensures that any set of up to $X$ colluding servers cannot learn anything about the database $\mathbf{W}$.
\begin{align}
    \text{[$X$-Security] } && I(\mathbf{W}; \mathbf{S}_{\mathcal{X}}) = 0, \forall \mathcal{X} \subset [1:N], |\mathcal{X}| \leq X.
\end{align}
The setting $X=0$ corresponds to replicated storage, where we set $\mathbf{S}_{n}=\mathbf{W}, \forall n\in[1:N]$. 

There are $M$ users. The user $m, m\in[1:M]$ specifies the index $\theta_m$ which is uniform over $[1:K_{m}]$. The $M$ users jointly want to retrieve the message $\mathbf{W}(\theta_{1}, \theta_{2}, \cdots, \theta_{M})$. The $m^{th}$ user must keep its\footnote{The use of `it' instead of `he/she' for users reflects the motivating example of $M$-factor authentication, where different users may in fact be different inanimate devices owned by the same person.} index private against collusion among any set of up to $T_{m}$ servers. Each user must also keep its index private against other users. 

To this end, we assume for each $m\in[1:M]$, user $m$ has its own private randomness $\mathcal{Z}_{m}$. Note that $\mathcal{Z}_{m}$ is used to guarantee user $m$'s $T_m$-privacy against any $T_m$ colluding servers. The $N$ servers share\footnote{We need common randomness  at the servers only to ensure perfect inter-user privacy, as in \eqref{eq:interuser}. Remarkably, \emph{almost-perfect} inter-user privacy can be guaranteed (for large messages) even without common randomness at servers (see Corollary \ref{cor:epsilon}).} common randomness $\widetilde{\mathcal{Z}}$ that is not available to the users.  The independence among these entities is formalized as follows.
\begin{align}
\begin{split}
    &H(\mathbf{S}_{[1:N]}, \widetilde{\mathcal{Z}}, (\theta_{m})_{m \in [1:M]}, (\mathcal{Z}_{m})_{m \in [1:M]})\\
    = &H(\mathbf{S}_{[1:N]}) + H(\widetilde{\mathcal{Z}}) + \sum_{m \in [1:M]}H(\theta_{m}) + \sum_{m \in [1:M]}H(\mathcal{Z}_{m}).
\end{split}
\end{align}
In order to retrieve the desired message, user $m$ generates $N$ queries $Q_{1}^{(m,\theta_{m})}$, $Q_{2}^{(m,\theta_{m})}$, $\cdots, Q_{N}^{(m,\theta_{m})}$ based on its index $\theta_{m}$ and its private randomness $\mathcal{Z}_{m}$. Specifically,
\begin{align}
    H(Q_{[1:N]}^{(m,\theta_{m})} | \theta_{m}, \mathcal{Z}_{m}) = 0, \forall m \in [1:M].
\end{align}
The corresponding queries from all $M$ users, $(Q_{n}^{(m,\theta_{m})})_{m \in [1:M]}$ are sent to the $n^{th}$ server, for all $n\in[1:N]$. Upon receiving the queries, the $n^{th}$ server generates its answer $A_{n}^{(\theta_{1}, \cdots, \theta_{M})}$  as a function of the queries, the stored information and the server-side common randomness.
\begin{align}
    H(A_{n}^{(\theta_{1}, \cdots, \theta_{M})} | \mathbf{S}_{n}, (Q_{n}^{(m,\theta_{m})})_{m \in [1:M]}, \widetilde{\mathcal{Z}}) = 0.
\end{align}
The privacy constraints consist of two parts. 
\begin{enumerate}
\item $(T_{m})$-Privacy. This means that  any $T_{m}$ or fewer servers have no knowledge about $\theta_m$,
\begin{align}
 I(\theta_{m}; (Q_{\mathcal{T}}^{(i, \theta_{i})})_{i \in [1:M]} | \mathbf{S}_{\mathcal{T}}, \widetilde{\mathcal{Z}}) = 0, && \forall m \in [1:M], \mathcal{T}\subset [1:N], |\mathcal{T}|\leq T_{m}.
\end{align}
\item Inter-user Privacy. This means that any user must learn nothing about other users' indices.
\begin{align}
    I\bigg(\theta_{[1:M]\backslash \{m\}}; A_{[1:N]}^{(\theta_{1}, \cdots, \theta_{M})} | \theta_{m}, \mathcal{Z}_{m}, \mathbf{W}(\theta_{1}, \cdots, \theta_{M})\bigg) = 0,&& \forall m \in [1:M].\label{eq:interuser}
\end{align}
\end{enumerate}
With  the answers from the $N$ servers, each user must be able to recover the desired message.
\begin{align}
    \text{[Correctness]} && 
     H(\mathbf{W}(\theta_{1}, \cdots, \theta_{M}) | A_{[1:N]}^{(\theta_{1},\cdots, \theta_{M})}, \theta_{m}, \mathcal{Z}_{m}) = 0, &&\forall m \in [1:M].
\end{align}
Recall that the rate of a PIR scheme is the number of bits of desired message that can be retrieved per bit of total download. Therefore, if $D$ is the maximum (over all realizations of messages) number of $q$-ary symbols downloaded from all servers by a user, under an MB-XS-TPIR scheme that allows the user to retrieve $L$ $q$-ary symbols of the desired message, then the rate of such a scheme is denoted as,
\begin{align}
    R = \frac{L}{D}.
\end{align}
The main contribution of this work is an achievable scheme for MB-XS-TPIR that is based on cross-subspace alignment, and achieves the rate $1-(X+T_1+\cdots+T_M)/N$, for arbitrary number of messages $K_1, K_2, \cdots, K_M$. Note that the scheme itself is not limited to asymptotic settings. Asymptotic settings will be of interest primarily for the purpose of testing the optimality of the scheme for significant special cases.

In order to introduce the scheme in a transparent setting, and to gain deeper insights into its optimality, we focus in particular on Double Blind $T$-PIR (DB-TPIR), which is obtained as a special case of MB-XS-TPIR by setting $M=2, X=0$. Given $q, L, N, K_1, K_2, T_1, T_2$ let us denote the supremum of rates achievable by any DB-TPIR scheme with these parameters as $R^*_{\mbox{\tiny DB-TPIR}}(q,L, N, K_1, K_2, T_1, T_2)$. Let us then define the capacity of DB-TPIR with parameters $N, K_1, K_2, T_1, T_2$ as
\begin{align}
    C_{\mbox{\tiny DB-TPIR}}(N, K_{1}, K_{2}, T_{1}, T_{2}) = \sup_{q,L}R^*_{\mbox{\tiny DB-TPIR}}(q,L, N, K_1, K_2, T_1, T_2).
\end{align}
Specifically, from the optimality perspective, we are interested in the asymptotic capacity of DB-TPIR as $K_1,K_2\rightarrow\infty$. Under the bounded latency $(b.l.)$ constraint, this asymptotic capacity is defined as
\begin{align}
    C^{\infty, b.l.}_{\mbox{\tiny DB-TPIR}}(N, T_{1}, T_{2})&\define \sup_{q,L}\lim_{K_1,K_2\rightarrow\infty}R^*_{\mbox{\tiny DB-TPIR}}(q,L, N, K_1, K_2, T_1, T_2).\label{eq:capbl}
\end{align}
In plain words,  $C^{\infty,b.l.}_{\mbox{\tiny DB-TPIR}}(N, T_{1}, T_{2})$ is the highest rate possible for any DB-TPIR scheme when the number of messages is much larger than the number of bits of each message that are jointly encoded by the scheme.

\begin{remark}
    For a double sequence $s(K_1,K_2)$, the notation $\lim_{K_1,K_2\rightarrow\infty}s(K_1,K_2)=a$ means that  $\forall\epsilon>0, \exists \kappa=\kappa(\epsilon)$ such that $|s(K_1,K_2)-a|<\epsilon, \forall K_1, K_2\geq \kappa$. (see Definition 2.1 in \cite{Habil_Double_Sequence}). It  follows from Theorem 4.2 in  \cite{Habil_Double_Sequence} that the double limit $\lim_{K_1,K_2\rightarrow\infty} R^*_{\mbox{\tiny DB-TPIR}}$ exists. This is because $R^*_{\mbox{\tiny DB-TPIR}}$ is a decreasing  sequence in each of $K_1$ and $K_2$ parameters individually (because any scheme that works with more messages also works with fewer messages), and is bounded below by zero. It also follows from Theorem 4.2 in  \cite{Habil_Double_Sequence} that  \begin{align}\lim_{K_1,K_2\rightarrow\infty} R^*_{\mbox{\tiny DB-TPIR}} = \lim_{K_1\rightarrow\infty}\lim_{K_2\rightarrow\infty}R^*_{\mbox{\tiny DB-TPIR}}.\label{eq:nolemma2}\end{align}
\end{remark}

\begin{remark}
    Note that the bounded-latency constraint affects the order in which the supremum is taken over message size parameters ($q,L$) versus the limit on the number of messages $(K_1, K_2)$. Without the bounded latency constraint,  the asymptotic capacity as the number of messages approaches infinity, would be defined as
\begin{align}
C^{\infty}_{\mbox{\tiny DB-TPIR}}(N, T_{1}, T_{2})&=\lim_{K_1,K_2\rightarrow\infty}\sup_{q,L}R^*_{\mbox{\tiny DB-TPIR}}(q,L, N, K_1, K_2, T_1, T_2)\notag\\
&= \lim_{K_1,K_2\rightarrow\infty}C_{\mbox{\tiny DB-TPIR}}(N, K_{1}, K_{2}, T_{1}, T_{2}).\label{eq:asymC}
\end{align}
\end{remark}

Comparing \eqref{eq:asymC} with \eqref{eq:capbl}, we note the key difference is that in \eqref{eq:asymC},  the supremum over message size ($q, L$)  allows message sizes to approach infinty for a fixed number of messages, and only then the number of messages approaches infinity, whereas in \eqref{eq:capbl} it is the number of messages $(K_1, K_2)$ that approaches infinity first for a given message size ($q, L$ are bounded, i.e., $O(1)$ in $K_1, K_2$), and only then the size of the message is allowed to grow. In a nutshell,  \eqref{eq:asymC} corresponds to asymptotic settings with $q^L\gg K_1,K_2$, while \eqref{eq:capbl} corresponds to asymptotic settings with $q^L\ll K_1,K_2$, thus prioritizing coding latency.

\section{Results}\label{sec: main}
We begin with the asymptotic capacity characterization of DB-TPIR under the bounded-latency constraint.

\begin{theorem}\label{thm: cap}
    The asymptotic capacity of DB-TPIR subject to bounded-latency constraint is 
    \begin{align}
    \begin{split}
        C^{\infty, b.l.}_{\mbox{\tiny DB-TPIR}}(N, T_{1}, T_{2}) =\left(1-\left(\frac{T_{1} + T_{2}}{N}\right)\right)^{+}.\\
    \end{split}\label{eq:cap}
    \end{align}
\end{theorem}
The proof of Theorem \ref{thm: cap} is presented in Section \ref{sec: DB-TPIR}. Notably, the achievability of the rate expression that appears on the RHS of \eqref{eq:cap}  needs neither the bounded-latency assumption, nor the asymptotic setting. Both of those are needed primarily for the converse argument.

Next we examine the need for common randomness across servers. Common randomness is needed across servers primarily to preserve inter-user privacy, i.e., to keep each user's index private from other users. While in the absence of common randomness, our achievable scheme does not  preserve inter-user privacy \emph{perfectly}, it is remarkable that the scheme manages to preserve inter-user privacy \emph{almost-perfectly} for large alphabet. In other words, the amount of information leaked to a user about the other user's index, is vanishingly small as $q\rightarrow\infty$.
Corollary \ref{cor:epsilon} highlights this observation by studying explicitly the case $T_1 = T_2 = 1, K_{1} = K_{2} = K$.
\begin{corollary}\label{cor:epsilon}
    For the DB-TPIR scheme proposed in Section \ref{sec: ach}, let $B_{[1:N]}^{(\theta_{1}, \theta_{2})}$ denote the answers generated by the $N$ servers after eliminating common randomness between servers (setting all symbols associated with $\widetilde{\mathcal{Z}}$ to zero in our achievable scheme for DB-TPIR). For $T_1 = T_2 = 1, K_{1} = K_{2} = K$ where $K$ is a fixed positive integer, and for any $\epsilon>0$, there exists $q_{0}>0$ s.t. when $q \geq q_{0}$ ($q$ is the size of the finite field $\mathbb{F}_{q}$),
    \begin{align}
       I(\theta_{2}; B_{[1:N]}^{(\theta_{1}, \theta_{2})} | \theta_{1}, \mathcal{Z}_{1}, \mathbf{W}(\theta_{1}, \theta_{2})) \leq \epsilon,\label{eq: epsilon1}\\
        I(\theta_{1}; B_{[1:N]}^{(\theta_{1}, \theta_{2})} | \theta_{2}, \mathcal{Z}_{2}, \mathbf{W}(\theta_{1}, \theta_{2})) \leq \epsilon.\label{eq: epsilon2}   
    \end{align}
\end{corollary}
The proof of Corollary \ref{cor:epsilon} appears in Appendix \ref{app:corepsilon}. 

Our final result generalizes the achievable scheme from DB-TPIR to MB-XS-TPIR based on a tensor-product extension of cross-subspace alignment. The achievable rate of the general scheme is presented in the following theorem.

\begin{theorem}\label{thm: rate}
    For the MB-XS-TPIR problem defined in Section \ref{sec: form}, the following rate is achievable regardless of the number of messages $K_1, K_2, \cdots, K_M$.
    \begin{align}
        R_{\mbox{\tiny MB-XS-TPIR}} = 1 - \frac{X + T_{1} + T_{2} + \cdots + T_{M}}{N}.
    \end{align}
\end{theorem}
Intuitively, this rate expression indicates that with this scheme one symbol is downloaded from each server, and from those $N$ symbols each user is able to recover $L=N-(X+T_1+T_2+\cdots+T_M)$ symbols of the desired message $\mathbf{W}(\theta_1,\theta_2,\cdots,\theta_M)$, while the interference is aligned within $X+T_1+T_2+\cdots+T_M$ dimensions. 
 Theorem \ref{thm: rate} is proved in Section \ref{sec: MB-XS-TPIR}.

\begin{corollary}\label{cor:XS-TPIR}
    Let us denote the supremum of achievable rates of MB-XS-TPIR (over all valid MB-XS-TPIR schemes) for fixed parameters $q,L,N,X,K_{1},\cdots,K_{M},T_{1},\cdots,T_{M}$ as $R^*_{\mbox{\tiny MB-XS-TPIR}}$. Further, let us define the capacity of MB-XS-TPIR as $C_{\mbox{\tiny MB-XS-TPIR}} = \sup_{q,L}R^*_{\mbox{\tiny MB-XS-TPIR}}$. Then we have the following bounds,
    \begin{align}
        1 - \frac{X + T_{1} + T_{2} + \cdots + T_{M}}{N} \leq C_{\mbox{\tiny MB-XS-TPIR}} \leq \min\bigg(\frac{1 - \frac{T_1+X}{N}}{1 - (\frac{T_{1}}{N-X})^{K_1}}, \cdots, \frac{1 - \frac{T_M+X}{N}}{1 - (\frac{T_{M}}{N-X})^{K_M}}\bigg).\label{cor:bound}
    \end{align}
\end{corollary}

The proof of Corollary \ref{cor:XS-TPIR} appears in Appendix \ref{app:corXS-TPIR}. The lower bound  in \eqref{cor:bound} follows directly from the proof of achievability of Theorem \ref{thm: rate}. The upper bound in \eqref{cor:bound} is obtained by noting that  MB-XS-TPIR schemes automatically yield XS-TPIR schemes. By setting $M = 2$ and $X = 0$, the capacity of DB-TPIR is bounded as
\begin{align}
    1 - \frac{T_{1}+T_{2}}{N} \leq C_{\mbox{\tiny DB-TPIR}}\leq \min\bigg(\frac{1 - T_1/N}{1 - (T_{1}/N)^{K_1}}, \frac{1 - T_{2}/N}{1 - (T_{2}/N)^{K_2}}\bigg).
\end{align}

\section{Asymptotic Capacity of DB-TPIR} \label{sec: DB-TPIR}
This section is devoted to the proof of Theorem \ref{thm: cap}. 

\subsection{Theorem \ref{thm: cap}: Converse} \label{sec: converse}

Let us find an upper bound on the capacity of DB-TPIR by noting a relationship between DB-TPIR and $X$-secure $T$-private information retrieval (XS-TPIR)  \cite{Jia_Sun_Jafar_XSTPIR}. Recall that XS-TPIR is a special case of MB-XS-TPIR obtained by setting $M=1$. The capacity of XS-TPIR with $N$ distributed servers, $K$ messages, $X$-secure  data storage, and $T$-private queries is denoted as $C_{\mbox{\tiny XS-TPIR}}(N, K, X, T)$. Recall that the asymptotic capacity of XS-TPIR (as $K\rightarrow\infty$) is shown in \cite{Jia_Sun_Jafar_XSTPIR} to be $C^\infty_{\mbox{\tiny XS-TPIR}}(N,X, T)=\left(1-\frac{X+T}{N}\right)^+$.

We will need the following lemma.

\begin{lemma}\label{lem: XS-TPIR1}
Let $R^*_{\mbox{\tiny DB-TPIR}}(q, L, N, K_{1}, K_{2}, T_{1}, T_{2})$ denote the supremum of rates achievable by any  DB-TPIR scheme for the  parameters $q, L, N, K_1, K_2, T_1, T_2$ as defined in Section \ref{sec: form}. Then for $K_2=q^{LK_1}$, we have,
\begin{align}
    R^*_{\mbox{\tiny DB-TPIR}}(q, L, N, K_{1}, K_{2} = q^{LK_{1}}, T_{1}, T_{2}) \leq C_{\mbox{\tiny XS-TPIR}}(N, K=K_{1}, X = T_{2}, T = T_{1}).\label{eq: cjlecx1}
\end{align}
\end{lemma}

\proof Consider a $K_1\times K_2$ matrix $\dot{\mathbf{W}}$ whose elements are from $\mathbb{F}_q^L$. The $K_2$ column vectors are all distinct and, say, arranged in lexicographic order. Since $K_2=q^{LK_1}$, the column vectors of the matrix  include all $q^{LK_1}$ possible realizations of $K_1\times 1$ vectors over $\mathbb{F}_q^L$, and $\dot{\mathbf{W}}$ is uniquely specified. We claim that any construction of a DB-TPIR scheme for the  parameter values specified on the LHS of \eqref{eq: cjlecx1}, when applied with the particular realization of the database $\mathbf{W}=\dot{\mathbf{W}}$, yields an XS-TPIR scheme with the parameters specified on the RHS of \eqref{eq: cjlecx1}.

Let us describe this XS-TPIR scheme. In this XS-TPIR scheme the user corresponds to User $1$ of the DB-TPIR scheme. Each Server $n$ stores only $Q_n^{(2,\theta_2)}$. Note that $\dot{\mathbf{W}}$ is a constant matrix known to everyone, whose $\theta_2^{th}$ column specifies the realizations of the $K_1$ i.i.d. messages (one of which is desired by the user), each comprised of $L$ uniformly random i.i.d. symbols from $\mathbb{F}_q$. Since $\theta_2$ is $T_2$-private according to the DB-TPIR construction, this constitutes $X=T_2$-secure storage of the $K_1$ messages. Furthermore, based on the $T_1$-private queries, $Q_n^{(1,\theta_1)}$, provided by the user, each server is able to respond as in the DB-TPIR scheme (because $Q_n^{(2,\theta_2)}$ is already known to Server $n$), and the DB-TPIR construction guarantees that the desired message $\dot{\mathbf{W}}(\theta_1,\theta_2)$ is correctly retrieved. Finally, since the rate of an XS-TPIR scheme cannot be higher than the capacity of XS-TPIR, the proof of Lemma  \ref{lem: XS-TPIR1} is complete.$\hfill\square$

\begin{remark}
    The XS-TPIR scheme that we obtain from the DB-TPIR scheme  described above, allows common randomness between servers. While the original formulation of XS-TPIR in \cite{Jia_Sun_Jafar_XSTPIR} does not explicitly allow common randomness, it is readily verified that server-side common randomness can be included in the storage of each server in the model of \cite{Jia_Sun_Jafar_XSTPIR}, and the asymptotic capacity result still holds.
\end{remark} 

\subsection*{Proof of Converse of Theorem \ref{thm: cap}}
Note that although the proof of Lemma \ref{lem: XS-TPIR1} requires the condition that $K_{1} = q^{LK_2}$, Theorem \ref{thm: cap} must hold as long as both $K_1$ and $K_2$ grow unbounded, regardless of their growth rates. For this we will utilize \eqref{eq:nolemma2}  as follows.
\begin{align}
C^{\infty, b.l.}_{\mbox{\tiny DB-TPIR}}(N, T_{1}, T_{2})&=\sup_{q,L}\lim_{K_1,K_2\rightarrow\infty}R^*_{\mbox{\tiny DB-TPIR}}(q,L, N, K_1, K_2, T_1, T_2)\\
&=\sup_{q,L}\lim_{K_{1}\rightarrow\infty}\bigg(\lim_{K_{2}\rightarrow\infty}R^*_{\mbox{\tiny DB-TPIR}}(q,L, N, K_1, K_2, T_1, T_2)\bigg)\label{eq:split}\\
&\leq \sup_{q,L}\lim_{K_{1}\rightarrow\infty}\bigg(\lim_{K_{2}\rightarrow\infty} R^*_{\mbox{\tiny DB-TPIR}}(q,L, N, \log_{q^L}(K_{1}), K_2, T_1, T_2)\bigg)\label{eq:reduce1}\\
&\leq \sup_{q,L}\lim_{K_{1}\rightarrow\infty} R^*_{\mbox{\tiny DB-TPIR}}(q,L, N, \log_{q^L}(K_{1}), K_1, T_1, T_2)\label{eq:reduce2}\\
&\leq \sup_{q,L}\lim_{K_{1}\rightarrow\infty} C_{\mbox{\tiny XS-TPIR}}(N, K=\log_{q^L}(K_{1}), X = T_{2}, T = T_{1}) \label{eq:uselemma}\\ 
&= \sup_{q,L}\lim_{K\rightarrow\infty} C_{\mbox{\tiny XS-TPIR}}(N, K, X = T_{2}, T = T_{1}) \label{eq:capnoL}\\ 
&= \lim_{K\rightarrow\infty} C_{\mbox{\tiny XS-TPIR}}(N, K, X = T_{2}, T = T_{1}) \label{eq:noL}\\
&    = 
    \left\{
    \begin{array}{ll}
    1-\left(\frac{T_{1} + T_{2}}{N}\right),& N> T_{1} + T_{2}\\
    0,& N\leq T_{1} + T_{2}.
    \end{array}
    \right.    \label{eq: cXS-TPIRlim}
\end{align}

The first step, \eqref{eq:split}, follows directly from  \eqref{eq:nolemma2}. In \eqref{eq:reduce1} we used the fact that reducing the number of messages cannot hurt the rate (because the original scheme can still be used with fewer messages). The next step, \eqref{eq:reduce2} follows because when $K_{2} \rightarrow\infty$, $K_{1}$ is viewed as a constant which is less than $K_{2}$ and reducing the number of messages cannot hurt the rate. For \eqref{eq:uselemma} we used Lemma \ref{lem: XS-TPIR1}. The next step, \eqref{eq:capnoL} follows  because for fixed $q,L$, and $K=\log_{q^L}(K_{1})$, the condition that $K_{1}\rightarrow\infty$ is equivalent to the condition that $K\rightarrow\infty$. Next, \eqref{eq:noL} follows because the capacity expression is not a function of $q$ or $L$. Finally, the asymptotic capacity characterization for XS-TPIR from \cite{Jia_Sun_Jafar_XSTPIR} is used for  \eqref{eq: cXS-TPIRlim}.
Thus, the proof of the converse part of Theorem \ref{thm: cap} is complete. $\hfill\square$

\subsection{Theorem \ref{thm: cap}: Achievability}\label{sec: ach}
In this section, we prove the achievability of Theorem \ref{thm: cap} by constructing  a scheme based on Cross Subspace Alignment (CSA) Codes \cite{Jia_Jafar_CDBC}, that can achieve the rate \\$\left(1 - (T_{1} + T_{2})/N\right)^+$ for arbitrary $N, K_{1}, K_{2}, T_{1}, T_{2}$. We will focus only on the non-trivial case, $N > T_{1} + T_{2}$. Throughout this scheme we  set,
\begin{align}
    L = N - (T_{1} + T_{2}).
\end{align}

Each message $\mathbf{W}(i,j), i \in [1:K_{1}], j \in [1:K_{2}]$ consists of $L$ symbols from finite field $\mathbb{F}_{q}$, denoted as $\mathbf{W}(i,j) = (\mathbf{W}(i,j)^{(1)}, \mathbf{W}(i,j)^{(2)}, \cdots, \mathbf{W}(i,j)^{(L)})$. For the scheme we will need the following $L+N$ distinct constants from $\mathbb{F}_q$,
\begin{align}
    f_{1}, f_{2}, \cdots, f_{L}, \alpha_{1}, \alpha_{2}, \cdots, \alpha_{N}
\end{align}
that are known to all  $N$ servers and the $2$ users. Note that this implies that $q\geq L+N$.

Let us split the messages $\mathbf{W}$ into $L$ matrices $(\mathbf{W}^{(1)}, \mathbf{W}^{(2)}, \cdots, \mathbf{W}^{(L)})$ so that $\mathbf{W}^{(l)}, l \in [1:L]$ contains the $l^{th}$ symbol of each message. Specifically,
\begin{align}
    \mathbf{W}^{(l)} = 
    \left[
    \begin{array}{cccc} 
      \mathbf{W}(1,1)^{(l)} & \mathbf{W}(1,2)^{(l)} & \cdots & \mathbf{W}(1,K_{2})^{(l)}\\
      \mathbf{W}(2,1)^{(l)} & \mathbf{W}(2,2)^{(l)} & \cdots & \mathbf{W}(2,K_{2})^{(l)}\\
      \vdots & \vdots & \vdots & \vdots\\
      \mathbf{W}(K_{1},1)^{(l)} & \mathbf{W}(K_{1},2)^{(l)} & \cdots & \mathbf{W}(K_{1},K_{2})^{(l)}\\
    \end{array}
    \right].
\end{align}
Note that we write equivalently $\mathbf{W}^{(l)}(\theta_{1}, \theta_{2}) = \mathbf{W}(\theta_{1}, \theta_{2})^{(l)}$.

Recall that $\mathbf{e}_{K}(\theta)$ is the $\theta^{th}$ column of the $K\times K$ identity matrix. The $l^{th}$ symbol of $\mathbf{W}(\theta_{1}, \theta_{2})$ can be  expressed as
\begin{align}
    \mathbf{W}(\theta_{1}, \theta_{2})^{(l)} = \mathbf{e}_{K_{1}}(\theta_1)^{\prime} \mathbf{W}^{(l)} \mathbf{e}_{K_{2}}(\theta_2).
\end{align}
Note here $\mathbf{e}_{K_{1}}(\theta_1)^{\prime} \mathbf{W}^{(l)}$ is the $\theta_{1}^{th}$ row of matrix $\mathbf{W}^{(l)}$. The inner product of $\theta_{1}^{th}$ row with $\mathbf{e}_{K_{2}}(\theta_2)$ is the entry at the $\theta_{2}^{th}$ column of this row, i.e., $\mathbf{W}(\theta_{1}, \theta_{2})^{(l)}$. The proposed scheme will enable the 2 users to retrieve $\mathbf{W}(\theta_{1}, \theta_{2})^{(l)}, \forall l \in [1:L]$, thus, retrieving $\mathbf{W}(\theta_{1}, \theta_{2})$.

The private randomness available to each user is specified as,
\begin{align}
    \mathcal{Z}_{1} = \{\mathbf{Z}_{1,t}^{(l)} \mid t \in [1:T_{1}], l \in [1:L]\},\\
    \mathcal{Z}_{2} = \{\mathbf{Z}_{2,t}^{(l)} \mid t \in [1:T_{2}], l \in [1:L]\}.
\end{align}
The random vectors $\mathbf{Z}_{1,t}^{(l)}\in\mathbb{F}_q^{K_1\times 1}, \mathbf{Z}_{2,t}^{(l)}\in\mathbb{F}_q^{K_2\times 1}$ have their elements drawn i.i.d. uniform from $\mathbb{F}_q$.

The query sent by user $m, m \in \{1,2\}$ to the $n^{th}$ server, $Q_{n}^{(m,\theta_{m})}$ is constructed as $Q_{n}^{(m,\theta_{m})} = (Q_{n,1}^{(m,\theta_{m})},Q_{n,2}^{(m,\theta_{m})},\cdots,Q_{n,L}^{(m,\theta_{m})})$ where $\forall l \in [1:L]$
\begin{align}
    Q_{n,l}^{(m,\theta_{m})} = \mathbf{e}_{K_{m}}(\theta_m) + \sum_{t \in [1:T_{m}]}(f_{l} - \alpha_{n})^{t}\mathbf{Z}_{m,t}^{(l)}.
\end{align}
Specifically, $Q_{n,l}^{(m,\theta_{m})}\in\mathbb{F}_q^{K_m\times 1}$ can be viewed as the query from user $m$ to request the $l^{th}$ symbol of the wanted message. The $T_m$-privacy constraint is satisfied since $Q_{n,l}^{(m,\theta_{m})}$ is the Shamir's secret sharing \cite{Shamir} of $\mathbf{e}_{K_m}(\theta_{m})$. Up to $T_m$ colluding servers can learn nothing about $\mathbf{e}_{K_m}(\theta_{m})$, thus, learning nothing about $\theta_{m}$.

Upon receiving queries from both users, the $n^{th}$ server computes an intermediate result
\begin{align}
    B_{n}^{(\theta_{1}, \theta_{2})} &= \sum_{l \in [1:L]}\frac{1}{f_{l} - \alpha_{n}} {Q_{n,l}^{(1,\theta_{1})}}^{\prime} \mathbf{W}^{(l)} Q_{n,l}^{(2,\theta_{2})}\label{eq: ach2u1}\\
    \begin{split}\label{eq: ach2u2}
    &= \frac{1}{f_{1} - \alpha_{n}} \underbrace{\mathbf{e}_{K_{1}}(\theta_1)^{\prime} \mathbf{W}^{(1)} \mathbf{e}_{K_{2}}(\theta_2)}_{\mathbf{W}(\theta_{1}, \theta_{2})^{(1)}} + \cdots + \frac{1}{f_{L} - \alpha_{n}} \underbrace{ \mathbf{e}_{K_{1}}(\theta_1)^{\prime} \mathbf{W}^{(L)} \mathbf{e}_{K_{2}}(\theta_2)}_{\mathbf{W}(\theta_{1}, \theta_{2})^{(L)}}\\
    &\quad\quad+ I_{0}  + \alpha_{n} I_{1} + \cdots + \alpha_{n}^{T_{1} + T_{2} - 1} I_{T_{1} + T_{2} - 1}.
    \end{split}
\end{align}
From \eqref{eq: ach2u1} to \eqref{eq: ach2u2}, distributive law is used. Note that \eqref{eq: ach2u2} can be viewed as a polynomial of $\alpha_{n}$. The coefficients of the first $L$ terms are the $L$ symbols of the desired message. $I_{i}, i \in [0:T_{1} + T_{2} -1]$ stands for the remaining (interference) terms that are generated by the product of the matrices in \eqref{eq: ach2u1}. The highest power of $\alpha_{n}$ is $T_{1} + T_{2} - 1$ and can be found from
\begin{align*}
    \sum_{l \in [1:L]} (f_{l} - \alpha_{n})^{T_{1} + T_{2} -1} {\mathbf{Z}_{1,T_{1}}^{(l)}}^{\prime} \mathbf{W}^{(l)} \mathbf{Z}_{2,T_{2}}^{(l)}.
\end{align*}

Note that the interference terms of \eqref{eq: ach2u2}, except the one of the highest order, may contain some information of the index specified by a user. For example, $I_{0}$ contains
\begin{align*}
    \frac{1}{f_{l} - \alpha_{n}} \mathbf{e}_{K_1}(\theta_{1})^{\prime} \mathbf{W}^{(l)} (f_{l} - \alpha_{n}) \mathbf{Z}_{2,1}^{(l)} = \mathbf{e}_{K_1}(\theta_{1})^{\prime} \mathbf{W}^{(l)} \mathbf{Z}_{2,1}^{(l)},
\end{align*}
which means that User $2$ may get some information about the index $\theta_{1}$ specified by User $1$ from the interference terms. To protect against this leakage of information, server $n$ will add noise drawn from the  common randomness that is shared by all servers. The common randomness shared among $N$ servers is specified as,
\begin{align}
    \widetilde{\mathcal{Z}} = \{\widetilde{Z}_{i} \mid i \in [0:T_{1} + T_{2} - 1]\},
\end{align}
where $(\widetilde{Z}_{i})_{i \in [0: T_{1} + T_{2} - 1]}$ are $T_{1} + T_{2}$ random variables that are i.i.d. uniform over $\mathbb{F}_{q}$. Server $n$ will add the polynomial
\begin{align}
    \widetilde{Z}(\alpha_{n}) = \widetilde{Z}_{0} + \alpha_{n} \widetilde{Z}_{1} + \cdots + \alpha_{n}^{T_{1} + T_{2} - 1} \widetilde{Z}_{T_{1} + T_{2} - 1} 
\end{align}
to the intermediate result $B_{n}^{(\theta_{1}, \theta_{2})}$ to generate its answer $A_{n}^{(\theta_{1}, \theta_{2})}$. This is the answer sent to both users.
\begin{align}
    A_{n}^{(\theta_{1}, \theta_{2})} &= B_{n}^{(\theta_{1}, \theta_{2})} + \widetilde{Z}(\alpha)\\
    \begin{split}
    &= \frac{1}{f_{1} - \alpha_{n}} \mathbf{W}(\theta_{1}, \theta_{2})^{(1)} + \cdots + \frac{1}{f_{L} - \alpha_{n}} \mathbf{W}(\theta_{1}, \theta_{2})^{(L)}\\
    &\quad\quad+ \underbrace{(I_{0} + \widetilde{Z}_{0})}_{J_{0}} + \cdots + \alpha_{n}^{T_{1} + T_{2} - 1}\underbrace{(I_{T_{1} + T_{2} - 1} + \widetilde{Z}_{T_{1} + T_{2} - 1})}_{J_{T_{1} + T_{2} - 1}}.
    \end{split}\label{eq: ach2u3}
\end{align}

\noindent Rewriting \eqref{eq: ach2u3} in matrix multiplication form, we have 
\begin{align}
    \begin{bmatrix}
    A_{1}^{(\theta_{1}, \theta_{2})}\\
    A_{2}^{(\theta_{1}, \theta_{2})}\\
    \vdots\\
    A_{N}^{(\theta_{1}, \theta_{2})}
    \end{bmatrix}&=
    \underbrace{
    \begin{bmatrix}
    \frac{1}{f_{1}-\alpha_{1}}&\frac{1}{f_{2}-\alpha_{1}}&\cdots&\frac{1}{f_{L}-\alpha_{1}}&1&\alpha_{1}&\cdots&\alpha_{1}^{T_{1} + T_{2} - 1}\\
    \frac{1}{f_{1}-\alpha_{2}}&\frac{1}{f_{2}-\alpha_{2}}&\cdots&\frac{1}{f_{L}-\alpha_{2}}&1&\alpha_{2}&\cdots&\alpha_{2}^{T_{1} + T_{2} - 1}\\
    \vdots&\vdots&\vdots&\vdots&\vdots&\vdots&\vdots&\vdots\\
    \frac{1}{f_{1}-\alpha_{N}}&\frac{1}{f_{2}-\alpha_{N}}&\cdots&\frac{1}{f_{L}-\alpha_{N}}&1&\alpha_{N}&\cdots&\alpha_{N}^{T_{1} + T_{2} - 1}\\
    \end{bmatrix}
    }_{\mathbf{C}}
    \begin{bmatrix}
    \mathbf{W}(\theta_{1}, \theta_{2})^{(1)}\\
    \mathbf{W}(\theta_{1}, \theta_{2})^{(2)}\\
    \vdots\\
    \mathbf{W}(\theta_{1}, \theta_{2})^{(L)}\\
    J_{0}\\
    \vdots\\
    J_{T_{1} + T_{2} - 1}
    \end{bmatrix}.\label{eq: mat_ach2u}
\end{align}
The matrix $\mathbf{C}$ is a Cauchy-Vandermonde matrix of size $N \times N$ since $N = L + T_{1} + T_{2}$. Since $f_{l}, l \in [1:L], \alpha_{n}, n \in [1:N]$ are $L + N$ distinct elements of $\mathbb{F}_{q}$, according to \cite{Gasca_Martinez_Muhlbach}, $\mathbf{C}$ is invertible in $\mathbb{F}_{q}$. Thus, the answers from all the $N$ servers form an invertible function of $\mathbf{W}(\theta_{1}, \theta_{2}), J_{0}, \cdots J_{T_{1} + T_{2} - 1}$. In other words, the correctness constraint is satisfied. 

Let us consider the inter-user privacy. Without loss of generality, let us consider User $1$. We have
\begin{align}
    &I(\theta_{2}; A_{[1:N]}^{(\theta_{1},\theta_{2})} | \theta_{1}, \mathcal{Z}_{1}, \mathbf{W}(\theta_{1}, \theta_{2}))\\
    &= I(\theta_{2}; \mathbf{W}(\theta_{1}, \theta_{2}), J_{[0: T_{1} + T_{2} -1]} | \theta_{1}, \mathcal{Z}_{1}, \mathbf{W}(\theta_{1}, \theta_{2}))\\
    &=I(\theta_{2}; J_{[0: T_{1} + T_{2} -1]} | \theta_{1}, \mathcal{Z}_{1}, \mathbf{W}(\theta_{1}, \theta_{2})) = 0. \label{eq: 2u0p}
\end{align}
\eqref{eq: 2u0p} comes from the fact that $J_{[0: T_{1} + T_{2} -1]}$ are protected by $T_{1} + T_{2}$ random symbols shared among servers, which are uniformly i.i.d. over $\mathbb{F}_{q}$ and are independent of all other terms in \eqref{eq: 2u0p}.

Finally, note that since $L=N-(T_1+T_2)$ symbols of the desired message are retrieved from a total of $N$ downloaded symbols from all $N$ servers, the rate of this scheme is $L/N=1-(T_1+T_2)/N$.

\subsection{Examples for Illustration}
\subsubsection{$L = 1, T_{1} = T_{2} = 1$ with $N = 3$ Servers}
Since $L = 1, T = 1$, we neglect the $l, t$ on superscripts or subscripts of all symbols. The queries from the 2 users are listed as follows.
\begin{align*}
\begin{array}{cc}
\hline
&\mbox{Server `$n$'} \\
\hline
Q_{n}^{(1,\theta_{1})} & \mathbf{e}_{K_1}(\theta_{1})+(f_{1} - \alpha_n)\mathbf{Z}_{1}\\
\hline
Q_{n}^{(2,\theta_{2})} & \mathbf{e}_{K_2}(\theta_{2})+(f_{1} - \alpha_n)\mathbf{Z}_{2}\\
\hline
\end{array}
\end{align*}
The intermediate result is computed as
\begin{align*}
    B_{n}^{(\theta_{1},\theta_{2})} &= \frac{1}{f_{1} - \alpha_{n}}{Q_{n}^{(1,\theta_{1})}}^{\prime} \mathbf{W} Q_{n}^{(2,\theta_{2})}\\
    &= \frac{1}{f_{1} - \alpha_{n}} \cdot \Big(\mathbf{e}_{K_1}(\theta_{1})^{\prime} + (f_{1} - \alpha_{n})\mathbf{Z}_{1}^{\prime}\Big) \cdot \mathbf{W} \cdot \Big(\mathbf{e}_{K_2}(\theta_{2}) + (f_{1} - \alpha_{n})\mathbf{Z}_{2}\Big)\\
    &= \frac{1}{f_{1} - \alpha_{n}} \mathbf{e}_{K_1}(\theta_{1})^{\prime} \mathbf{W} \mathbf{e}_{K_2}(\theta_{2}) + \Big(\mathbf{Z}_{1}^{\prime} \mathbf{W} \mathbf{e}_{K_2}(\theta_{2}) + \mathbf{e}_{K_1}(\theta_{1})^{\prime} \mathbf{W} \mathbf{Z}_{2}\Big) + (f_{1} - \alpha_{n})\mathbf{Z}_{1}^{\prime} \mathbf{W} \mathbf{Z}_{2}\\
    &= \frac{1}{f_{1} - \alpha_{n}} \mathbf{W}(\theta_{1}, \theta_{2}) + \underbrace{ \Big(\mathbf{Z}_{1}^{\prime} \mathbf{W} \mathbf{e}_{K_2}(\theta_{2}) + \mathbf{e}_{K_1}(\theta_{1})^{\prime} \mathbf{W} \mathbf{Z}_{2} + f\mathbf{Z}_{1}^{\prime} \mathbf{W} \mathbf{Z}_{2}\Big)}_{I_{0}} + \alpha_{n}\underbrace{(-\mathbf{Z}_{1}^{\prime} \mathbf{W} \mathbf{Z}_{2})}_{I_{1}}.
\end{align*}
The answer from the server is
\begin{align*}
    A_{n}^{(\theta_{1}, \theta_{2})} &= B_{n}^{(\theta_{1}, \theta_{2})} + \widetilde{Z}_{0} + \alpha_{n}\widetilde{Z}_{1}\\
    &= \frac{1}{f_{1} - \alpha_{n}} \mathbf{W}(\theta_{1}, \theta_{2}) + J_{0} + \alpha_{n} J_{1}.
\end{align*}
Writing in matrix form, the answers from $N=3$ servers are
\begin{align*}
    \begin{bmatrix}
    A_{1}^{(\theta_{1}, \theta_{2})}\\
    A_{2}^{(\theta_{1}, \theta_{2})}\\
    A_{3}^{(\theta_{1}, \theta_{2})}
    \end{bmatrix}&=
    \underbrace{
    \begin{bmatrix}
    \frac{1}{f_{1}-\alpha_{1}}&1&\alpha_{1}\\
    \frac{1}{f_{1}-\alpha_{2}}&1&\alpha_{2}\\
    \frac{1}{f_{1}-\alpha_{3}}&1&\alpha_{3}\\
    \end{bmatrix}
    }_{\mathbf{C}}
    \begin{bmatrix}
    \mathbf{W}(\theta_{1}, \theta_{2})\\
    J_{0}\\
    J_{1}
    \end{bmatrix}.
\end{align*}
The desired message is retrieved by inverting the matrix $\mathbf{C}$. Since $L=N-(T_1+T_2)=1$ symbol of the desired message is retrieved from a total of $N=3$ downloaded symbols from all $3$ servers, the rate of the scheme is $L/N=1/3$.

\subsubsection{$L = 2, T_{1} = 1, T_{2} = 2$ with $N = 5$ Servers}
The queries from the 2 users are listed as follows.
\begin{align*}
\begin{array}{cc}
\hline
&\mbox{Server `$n$'} \\
\hline
Q_{n,1}^{(1,\theta_{1})} & \mathbf{e}_{K_1}(\theta_{1})+(f_{1} - \alpha_n)\mathbf{Z}_{1,1}^{(1)}\\
Q_{n,2}^{(1,\theta_{1})} & \mathbf{e}_{K_1}(\theta_{1})+(f_{2} - \alpha_n)\mathbf{Z}_{1,1}^{(2)}\\
\hline
Q_{n,1}^{(2,\theta_{2})} & \mathbf{e}_{K_2}(\theta_{2})+(f_{1} - \alpha_n)\mathbf{Z}_{2,1}^{(1)} + (f_{1} - \alpha_n)^{2}\mathbf{Z}_{2,2}^{(1)}\\
Q_{n,2}^{(2,\theta_{2})} & \mathbf{e}_{K_2}(\theta_{2})+(f_{2} - \alpha_n)\mathbf{Z}_{2,1}^{(2)} + (f_{2} - \alpha_n)^{2}\mathbf{Z}_{2,2}^{(2)}\\
\hline
\end{array}
\end{align*}
The intermediate result is
\begin{align*}
    B_{n}^{(\theta_{1},\theta_{2})} &= \frac{1}{f_{1} - \alpha_{n}}{Q_{n,1}^{(1,\theta_{1})}}^{\prime} \mathbf{W}^{(1)} Q_{n,1}^{(2,\theta_{2})} + \frac{1}{f_{2} - \alpha_{n}}{Q_{n,2}^{(1,\theta_{1})}}^{\prime} \mathbf{W}^{(2)} Q_{n,2}^{(2,\theta_{2})}\\
    & = \frac{1}{f_{1} - \alpha_{n}}\mathbf{W}(\theta_{1}, \theta_{2})^{(1)} + \frac{1}{f_{2} - \alpha_{n}}\mathbf{W}(\theta_{1}, \theta_{2})^{(2)} + I_{0} + \cdots + \alpha_{n}^{2}I_{2}.
\end{align*}
The answer is
\begin{align*}
    A_{n}^{(\theta_{1},\theta_{2})} = \frac{1}{f_{1} - \alpha_{n}}\mathbf{W}(\theta_{1}, \theta_{2})^{(1)} + \frac{1}{f_{2} - \alpha_{n}}\mathbf{W}(\theta_{1}, \theta_{2})^{(2)} + \underbrace{(I_{0} + \widetilde{Z}_{0})}_{J_{0}} + \cdots + \alpha_{n}^{2}\underbrace{(I_{2}+\widetilde{Z}_{2})}_{J_{2}}.
\end{align*}
Writing in matrix form, the answers from $N=5$ servers are
\begin{align*}
    \begin{bmatrix}
    A_{1}^{(\theta_{1}, \theta_{2})}\\
    A_{2}^{(\theta_{1}, \theta_{2})}\\
    A_{3}^{(\theta_{1}, \theta_{2})}\\
    A_{4}^{(\theta_{1}, \theta_{2})}\\
    A_{5}^{(\theta_{1}, \theta_{2})}
    \end{bmatrix}&=
    \underbrace{
    \begin{bmatrix}
    \frac{1}{f_{1}-\alpha_{1}}&\frac{1}{f_{2}-\alpha_{1}}&1&\alpha_{1}&\alpha_{1}^{2}\\
    \frac{1}{f_{1}-\alpha_{2}}&\frac{1}{f_{2}-\alpha_{2}}&1&\alpha_{2}&\alpha_{2}^{2}\\
    \frac{1}{f_{1}-\alpha_{3}}&\frac{1}{f_{2}-\alpha_{3}}&1&\alpha_{3}&\alpha_{3}^{2}\\
    \frac{1}{f_{1}-\alpha_{4}}&\frac{1}{f_{2}-\alpha_{4}}&1&\alpha_{4}&\alpha_{4}^{2}\\
    \frac{1}{f_{1}-\alpha_{5}}&\frac{1}{f_{2}-\alpha_{5}}&1&\alpha_{5}&\alpha_{5}^{2}
    \end{bmatrix}
    }_{\mathbf{C}}
    \begin{bmatrix}
    \mathbf{W}(\theta_{1}, \theta_{2})^{(1)}\\
    \mathbf{W}(\theta_{1}, \theta_{2})^{(2)}\\
    J_{0}\\
    J_{1}\\
    J_{2}
    \end{bmatrix}.
\end{align*}
Evidently, the rate achieved is $L/N=2/5$ in this case.

\section{$M$-way Blind $X$-Secure $T$-Private Information Retrieval}\label{sec: MB-XS-TPIR}
In this section, we propose a scheme that solves the generalized problem: $M$-way blind $X$-secure $T$-private information retrieval (MB-XS-TPIR). The rate achieved by this scheme is $R = 1 - (X + T_{1} + \cdots + T_{M})/N$.

MB-XS-TPIR has been formalized in Section \ref{sec: form}. In brief, MB-XS-TPIR enables $M$ users who independently specify $M$ indices $\theta_{1}, \cdots, \theta_{M}$ ($\theta_{m}$ is specified by user $m$) to retrieve a message $\mathbf{W}(\theta_{1}, \cdots, \theta_{M})$ from a database $\mathbf{W}$ which is $X$-securely stored at $N$ distributed servers, with $(T_{m})$-Privacy and User-User Privacy constraints satisfied.

The MB-XS-TPIR scheme proposed in this section is still based on Cross Subspace Alignment (CSA) and is a natural extension of the DB-TPIR scheme. The main difference is that in this case, the database $\mathbf{W}$  is an $M$-dimensional tensor instead of a 2-dimensional matrix in DB-TPIR. 

\subsection{Brief Review of Tensors}
Let us briefly review the key properties of tensors that we will need. Specifically, an $M$-dimensional tensor is an $M$-dimensional array. For instance a $2$-dimensional tensor is a matrix, and a $3$-dimensional tensor is a cuboid made up of several matrices. Each dimension of a tensor is called a mode. The $m^{th}$ dimension is called mode-$m$. The tensor operation we mainly need is the operation called \textit{mode-m tensor vector multiplication}. Readers can refer to Chapter 3, Section 3.1.2 of \cite{Lu_Platanioties_Venetsanopoulos_Tensor} for more details. 

\begin{definition}{\textbf{Mode-$m$ Tensor Vector Multiplication.}}
  The mode-$m$ multiplication of a tensor $\mathbf{A} \in \mathbb{F}_{q}^{K_{1} \times K_{2} \times \cdots \times K_{M}}$ with a column vector   $\mathbf{b} \in \mathbb{F}_{q}^{K_{m} \times 1}$ 
  results in the tensor,
\begin{align}
    \mathbf{C} = \mathbf{A} \times_{m} \mathbf{b},\label{eq:tvm1}
  \end{align}
  where $\mathbf{C} \in \mathbb{F}_{q}^{K_{1} \times \cdots \times K_{m-1} \times 1 \times K_{m+1} \times \cdots \times K_{M}}$, and  each element of $\mathbf{C}$ is specified as
  \begin{align}
    \mathbf{C}(k_{1}, \cdots, k_{m-1}, 1, k_{m+1}, \cdots, k_{M}) = \sum_{k_{m} \in [1: K_{m}]} \mathbf{A}(k_{1}, \ldots, k_{M}) \cdot \mathbf{b}(k_{m}).\label{eq:tvm2}
  \end{align}
\end{definition}

\noindent Note that this operation is a multilinear operation, so distributive law applies to this operation.

\subsection{General MB-XS-TPIR Scheme}\label{sec:genscheme}
Before formally presenting our MB-XS-TPIR solution, let us briefly explain at a high level how our solution  translates into the problem of secure distributed tensor product computation. For our  solution, we first arrange the data into $L$  tensors $\mathbf{W}^{(1)}, \cdots, \mathbf{W}^{(L)}$, where ${\bf W}^{(l)}\in\mathbb{F}_q^{K_1\times K_2\cdots \times K_M}, l\in[1:L]$ is comprised of the $l^{th}$ symbol of each of the $K_1 K_2\cdots K_M$ messages. The tensorized data is secret shared among the $N$ servers as $(\mathbf{S}_n^{(1)}, \cdots, \mathbf{S}_n^{(L)})_{n \in [1:N]}$ to guarantee  $X$-security. Next, the $M$ vectors $\mathbf{e}_{K_1}(\theta_1), \cdots, \mathbf{e}_{K_M}(\theta_M)$, corresponding to the indices specified by the $M$ users, are secret-shared among the $N$ servers in the form of the queries $(Q_{n}^{(1,\theta_{1})}, \cdots, Q_{n}^{(M,\theta_{M})})_{n \in [1:N]}$ to retrieve the desired message. $(Q_{n}^{(m,\theta_{m})})_{n\in[1:N]}$ is the secret-sharing of the query from the $m^{th}$ user that ensures $T_m$ privacy. Most importantly, with this construction of queries and tensorized data, retrieving the desired message  corresponds to retrieving tensor products of the privatized queries and secured data. From this point on, the achievability scheme for MB-XS-TPIR can indeed be viewed as a secure coded tensor product computation, which is an multilinear operation with $M+1$ inputs, for which CSA codes \cite{Jia_Jafar_CDBC} can be used. To optimize the download cost for MB-XS-TPIR, the parameters of the CSA codes are chosen as: $K_c=1, \ell=N-(X+\sum_{m\in[1:M]} T_m)$. Note that the proposed scheme automatically recovers  asymptotically optimal schemes for various special cases of MB-XS-TPIR, such as PIR, TPIR, XS-TPIR, etc. This further underscores the connection between various forms of PIR and coded distributed computing.

Now let us proceed to formally present our MB-XS-TPIR scheme. Throughout this scheme we set $L = N - (T_{1} + T_{2} + \cdots +T_{M}) - X$. Let $\mathbb{F}_{q}$ be a finite field with $q \ge L + N$ and let $f_{1}, \cdots, f_{L}, \alpha_{1}, \cdots, \alpha_{N}$ be $L + N$ distinct elements in $\mathbb{F}_{q}$. These $L + N$ elements are known to the $N$ servers and $M$ users. 

The private randomness available at user $m$ to keep its index $\theta_{m}$ $T_{m}$-private is
\begin{align}
    \mathcal{Z}_{m} = \{\mathbf{Z}_{m,t}^{(l)} \mid t \in [1:T_{m}], l \in [1:L]\},&& \forall m \in [1:M],
\end{align}
where the column vectors $\mathbf{Z}_{m,t}^{(l)}\in\mathbb{F}_q^{K_m\times 1}$ have entries drawn i.i.d. uniform from $\mathbb{F}_{q}$. 

For compact notation, we write $\sum T_m$ instead of $\sum_{m\in[1:M]} T_m$. The common randomness $\widetilde{\mathcal{Z}}$ shared among $N$ servers for protecting inter-user privacy is specified as 
\begin{align}
    \widetilde{\mathcal{Z}} = \left\{\widetilde{Z}_{i} \mid i \in \left[0: \sum T_{m} + X - 1\right]\right\},
\end{align}
where $\widetilde{Z}_{i}, i \in [0: \sum T_{m} + X - 1]$ are $\sum T_{m} + X$ random noise variables that are i.i.d. and uniform over $\mathbb{F}_{q}$.

To form $X$-secure storage of the data, let us introduce
\begin{align}
    \widehat{\mathcal{Z}} = \{\widehat{\mathbf{Z}}_{l,x} \mid x \in [1:X], l \in [1:L]\},
\end{align}
which are independent uniform random noise tensors from $\mathbb{F}_{q}^{K_{1} \times \cdots \times K_{M}}$. 

The database $\mathbf{W}$ can be split into $L$ parts, each of which is an $M$-dimensional tensor. This partitioning is specified as
\begin{align}
    \mathbf{W} = (\mathbf{W}^{(1)}, \mathbf{W}^{(2)}, \cdots, \mathbf{W}^{(L)}),&& \mathbf{W}^{(l)} \in \mathbb{F}_{q}^{K_{1} \times K_{2} \times \cdots \times K_{M}}, \forall l \in [1:L],
\end{align}
so that $\mathbf{W}^{(l)}$ contains the $l^{th}$ symbol of every message. 

The independence between the messages, indices, and noises is specified as
\begin{align}
    \begin{split}
        &H(\mathbf{W}, (\theta_{m})_{m \in [1:M]}, (\mathcal{Z}_{m})_{m \in[1:M]}, \widetilde{\mathcal{Z}}, \widehat{\mathcal{Z}})\\
        = &\sum_{l\in[1:L]} H(\mathbf{W}^{(l)}) + \sum_{m\in[1:M]} H(\theta_{m}) + \sum_{m\in[1:M]} H(\mathcal{Z}_{m}) + H(\widetilde{\mathcal{Z}}) + H(\widehat{\mathcal{Z}})\\
        = &LK_{1}\cdots K_{M} + \sum_{m\in[1:M]} H(\theta_{m}) + \sum_{m\in[1:M]} LK_{m}T_{m} + \sum_{m\in[1:M]}T_{m} + X + LK_{1}\cdots K_{M}X.
    \end{split}
\end{align}

To keep the database $\mathbf{W}$ $X$-secure, $\mathbf{W}$ is secret-shared among $N$ servers. The $n^{th}$ server holds the share $\mathbf{S}_{n} = (\mathbf{S}_{n}^{(1)}, \cdots, \mathbf{S}_{n}^{(L)})$ where
\begin{align}
    \mathbf{S}_{n}^{(l)} = \mathbf{W}^{(l)} + \sum_{x \in [1:X]}(f_{l} - \alpha_{n})^{x} \widehat{\mathbf{Z}}_{l,x}.
\end{align}

Note that $\mathbf{e}_{K}(\theta)$ is the $\theta^{th}$ column of the $K\times K$ identity matrix. With the tensor vector multiplication defined above, the desired message can be written as
\begin{align}
\begin{split}
    \mathbf{W}(\theta_{1}, \cdots, \theta_{M}) &= (\mathbf{W}^{(l)}(\theta_{1}, \cdots, \theta_{M}))_{l \in [1:L]}\\
    &=(\mathbf{W}^{(l)} \times_{1} \mathbf{e}_{K_1}(\theta_{1}) \times_{2} \mathbf{e}_{K_2}(\theta_{2}) \times_{3} \cdots \times_{M} \mathbf{e}_{K_M}(\theta_{M}))_{l \in [1:L]}.
\end{split}
\end{align}

To guarantee $T_{m}$-privacy, the index specified by the $m^{th}$ user is protected by $T_{m}$ random noise vectors. The queries sent from the $m^{th}$ user to the $n^{th}$ server are constructed as $Q_{n}^{(m,\theta_{m})} = (Q_{n,1}^{(m,\theta_{m})}, Q_{n,2}^{(m,\theta_{m})}, \cdots, Q_{n,L}^{(m,\theta_{m})})$ where 
\begin{align}
\begin{split}
    Q_{n,l}^{(m,\theta_{m})} = \mathbf{e}_{K_m}(\theta_{m}) + \sum_{t \in [1:T_{m}]}(f_{l} - \alpha_{n})^{t}\mathbf{Z}_{m,t}^{(l)}, \forall l \in [1:L], m \in [1:M].
\end{split}    
\end{align}

With the queries from the $M$ users and stored $\mathbf{S}_{n}$, the $n^{th}$ server first computes an intermediate result 
\begin{align}
\begin{split}
    B_{n}^{(\theta_{1}, \theta_{2}, \cdots, \theta_{M})} &= \sum_{l \in [1:L]} \frac{1}{f_{l} - \alpha_{n}} \mathbf{S}_{n}^{(l)} \times_{1} Q_{n,l}^{(1,\theta_{1})} \times_{2} Q_{n,l}^{(2\theta_{2})} \times_{3} \cdots \times_{M} Q_{n,l}^{(M,\theta_{M})}\\
    &= \sum_{l \in [1:L]} \frac{1}{f_{l} - \alpha_{n}} \mathbf{W}^{(l)} \times_{1} \mathbf{e}_{K_1}(\theta_{1}) \times_{2} \cdots \times_{M} \mathbf{e}_{K_M}(\theta_{M}) + I_{0} + \alpha_{n} I_{1} + \cdots \\
    &\quad\quad+ \alpha_{n}^{\sum T_{m} + X - 1} I_{\sum T_{m} + X - 1}\\
    &= \sum_{l \in [1:L]} \frac{1}{f_{l} - \alpha_{n}}\mathbf{W}^{(l)}(\theta_{1}, \cdots, \theta_{M}) + I_{0} + \alpha_{n} I_{1} + \cdots + \alpha_{n}^{\sum T_{m} + X - 1} I_{\sum T_{m} + X - 1}.
\end{split}
\end{align}

As before, $I_{0}, \cdots, I_{\sum T_{m} + X - 1}$ are $\sum T_{m} + X$ interference terms which are useless. Note that the distributive law applies here because mode-$m$ multiplication is a multilinear operation. The highest order of $\alpha_{n}$ is $\sum T_{m} + X - 1$, which results from 
\begin{align}
    \sum_{l \in [1:L]} \frac{1}{f_{l} - \alpha_{n}}(f_{l} - \alpha_{n})^{X} \widehat{\mathbf{Z}}_{l,X} \times_{1} (f_{l} - \alpha_{n})^{T_{1}} \mathbf{Z}_{1, T_{1}}^{(l)}  \times_{2} \cdots \times_{M} (f_{l} - \alpha_{n})^{T_{M}} \mathbf{Z}_{M, T_{M}}^{(l)}.
\end{align}

Similar to DB-TPIR, the interference terms may contain some information of the indices specified by all users. To guarantee privacy between users, servers will add common randomness shared among them to the intermediate results to generate their answers for each user. Specifically, the answer from server $n$ is  
\begin{align}
\begin{split} \label{eq: achMB-XS-TPIR}
    A_{n}^{(\theta_{1}, \cdots, \theta_{M})} &= B_{n}^{(\theta_{1}, \cdots, \theta_{M})} + \widetilde{Z}_{0} + \alpha_{n} \widetilde{Z}_{1} + \cdots + \alpha_{n}^{\sum T_{m} + X - 1} \widetilde{Z}_{\sum T_{m} + X - 1}\\
    &= \sum_{l \in [1:L]} \frac{1}{f_{l} - \alpha_{n}}\mathbf{W}^{(l)}(\theta_{1}, \cdots, \theta_{M}) + \underbrace{(I_{0} + \widetilde{Z}_{0})}_{J_{0}} + \cdots \\
    &\quad\quad+ \alpha_{n}^{\sum T_{m} + X - 1} \underbrace{(I_{\sum T_{m} + X - 1} + \widetilde{Z}_{\sum T_{m} + X - 1})}_{J_{\sum T_{m} + X - 1}}.
\end{split}
\end{align}

The matrix form of \eqref{eq: achMB-XS-TPIR} is similar to \eqref{eq: mat_ach2u}, we omit it here. Since $L = N - \sum T_{m} - X$ dimensions are occupied by desired message symbols and $\sum T_{m} + X$ dimensions are occupied by the noisy versions of interference terms ($J$), the rate achieved here is 
\begin{align}
    R = \frac{L}{N} = 1 - \frac{\sum T_{m} + X}{N}.
\end{align}

\subsection{Example}
Let us provide a simple example for illustration.

\textit{$N = 8$ Servers, $M = 3$ users with $T_{1} = T_{2} = 1, T_{3} = 2$, $X = 2$, $L = 2$.}

The storage at Server $n$ and the queries from the $3$ users are listed as follows.
\begin{align*}
\begin{array}{cc}
\hline
&\mbox{Server `$n$'} \\
\hline
\mathbf{S}_{n}^{(1)} & \mathbf{W}^{(1)} + (f_{1} - \alpha_{n}) \widehat{\mathbf{Z}}_{1,1} + (f_{1} - \alpha_{n})^{2} \widehat{\mathbf{Z}}_{1,2}\\
\mathbf{S}_{n}^{(2)} & \mathbf{W}^{(2)} + (f_{2} - \alpha_{n}) \widehat{\mathbf{Z}}_{2,1} + (f_{2} - \alpha_{n})^{2} \widehat{\mathbf{Z}}_{2,2}\\
\hline
Q_{n,1}^{(1,\theta_{1})} & \mathbf{e}_{K_1}(\theta_{1})+(f_{1} - \alpha_{n})\mathbf{Z}_{1,1}^{(1)}\\
Q_{n,2}^{(1,\theta_{1})} & \mathbf{e}_{K_1}(\theta_{1})+(f_{2} - \alpha_{n})\mathbf{Z}_{1,1}^{(2)}\\
\hline
Q_{n,1}^{(2,\theta_{2})} & \mathbf{e}_{K_2}(\theta_{2})+(f_{1} - \alpha_{n})\mathbf{Z}_{2,1}^{(1)}\\
Q_{n,2}^{(2,\theta_{2})} & \mathbf{e}_{K_2}(\theta_{2})+(f_{2} - \alpha_{n})\mathbf{Z}_{2,1}^{(2)}\\
\hline
Q_{n,1}^{(3,\theta_{3})} & \mathbf{e}_{K_3}(\theta_{3})+(f_{1} - \alpha_{n})\mathbf{Z}_{3,1}^{(1)} + (f_{1} - \alpha_{n})^{2}\mathbf{Z}_{3,2}^{(1)}\\
Q_{n,2}^{(3,\theta_{3})} & \mathbf{e}_{K_3}(\theta_{3})+(f_{2} - \alpha_{n})\mathbf{Z}_{3,1}^{(2)} + (f_{2} - \alpha_{n})^{2}\mathbf{Z}_{3,2}^{(2)}\\
\hline
\end{array}
\end{align*}

The intermediate result is 
\begin{align*}
    B_{n}^{(\theta_{1}, \theta_{2}, \theta_{3})} = \frac{1}{f_{1} - \alpha_{n}} \mathbf{W}^{(1)}(\theta_{1}, \theta_{2}, \theta_{3}) + \frac{1}{f_{2} - \alpha_{n}}\mathbf{W}^{(1)}(\theta_{1}, \theta_{2}, \theta_{3}) + I_{0} + \cdots + \alpha_{n}^{5} I_{5}.
\end{align*}
The highest order of $\alpha$ is $5$ since $T_{1} + T_{2} + T_{3} + X - 1 = 5$ in this case. The answer from the server is 
\begin{align*}
    A_{n}^{(\theta_{1}, \theta_{2}, \theta_{3})} = \frac{1}{f_{1} - \alpha_{n}} \mathbf{W}^{(1)}(\theta_{1}, \theta_{2}, \theta_{3}) + \frac{1}{f_{2} - \alpha_{n}}\mathbf{W}^{(2)}(\theta_{1}, \theta_{2}, \theta_{3}) + \underbrace{(I_{0} + \widetilde{Z}_{0})}_{J_{0}} + \cdots + \alpha_{n}^{5} \underbrace{(I_{5} + \widetilde{Z}_{5})}_{J_{5}}.
\end{align*}
Evidently, the desired symbols occupy $2$ dimensions,  the aligned interference occupies $6$ dimensions, and the rate achieved is $2/8 = 1/4$.

To further explain the example intuitively, $(\mathbf{S}_{n}^{(l)})_{l\in[1:2]}$ can be viewed as the secret shares of $\mathbf{W}^{(1)}, \mathbf{W}^{(2)}$ for the $N$ servers, and $Q_{n,1}^{(1,\theta_{1})},Q_{n,2}^{(1,\theta_{1})}$ can be viewed as two independent shares of $\mathbf{e}_{K_1}(\theta_1)$ at the $n^{th}$ server, $n\in[1:N]$. Similarly, $Q_{n,1}^{(2,\theta_{2})},Q_{n,2}^{(2,\theta_{2})}$ and $Q_{n,1}^{(3,\theta_{3})},Q_{n,2}^{(3,\theta_{3})}$ are independent shares of $\mathbf{e}_{K_2}(\theta_2)$ and $\mathbf{e}_{K_3}(\theta_3)$, respectively. $B_n^{(\theta_1,\theta_2,\theta_3)}$ is constructed following the idea of CSA codes \cite{Jia_Jafar_CDBC} such that the interference symbols align within the $6$ dimensions of the subspace spanned by the Vandermonde terms, while the two desired symbols, represented as $\mathbf{W}^{(1)}(\theta_1,\theta_2,\theta_3)=\mathbf{W}^{(1)}\times_{1} Q_{n,l}^{(1,\theta_{1})} \times_{2} Q_{n,l}^{(2\theta_{2})} \times_{3} Q_{n,l}^{(3,\theta_{3})}$ and $\mathbf{W}^{(2)}(\theta_1,\theta_2,\theta_3)=\mathbf{W}^{(2)}\times_{1} Q_{n,l}^{(1,\theta_{1})} \times_{2} Q_{n,l}^{(2\theta_{2})} \times_{3} Q_{n,l}^{(3,\theta_{3})}$, remain resolvable along the Cauchy terms.

\section{Conclusion}\label{sec: conclusion}
We explored the problem of  $M$-way blind $X$-secure $T$-private information retrieval (MB-XS-TPIR). We found the asymptotic capacity of double blind $T$-private information retrieval (DB-TPIR), which is a special case of MB-XS-TPIR, under a bounded-latency constraint. The achievable scheme was constructed based on Cross-Subspace Alignment. We then generalized the scheme using tensor-products into an MB-XS-TPIR scheme where the number of users ($M$), storage security-level ($X$) and privacy level of each user's index ($T_1,T_2,\cdots,T_M$) can be arbitrarily chosen. 

This work leads to a number of open problems. Foremost is the question of optimality of the proposed solutions. For example, the asymptotic capacity for MB-XS-TPIR remains open. For non-asymptotic settings, the capacity remains open even for DB-TPIR. As discussed in the introduction, we expect that our solution to MB-XS-TPIR may be asymptotically optimal. In fact, we expect that our solution may be optimal even in non-asymptotic settings. This is because of the constraint that the user must learn nothing  about the other users' indices, which is reminiscent of `symmetric' privacy constraints in PIR. Prior works, e.g., \cite{Sun_Jafar_SPIR, Wang_Skoglund, Wang_Skoglund_SPIRAd, Wang_Sun_Skoglund_BSPIR}, suggest that the capacity of PIR under symmetric privacy constraints tends to be the same as the asymptotic capacity without symmetric privacy constraints. Another open problem is to characterize the minimal amount of common randomness needed to be shared among servers for MB-XS-TPIR. Finally, yet another promising direction for future work is the setting of secure multiparty computation where the messages ${\bf W}(\theta_1, \theta_2, \cdots, \theta_M)$ are deterministic functions of $(\theta_1, \theta_2, \cdots, \theta_M)$. What makes these settings  challenging is that their upload costs may not be negligible relative to download costs, so instead of a capacity figure the optimal solution may be a tradeoff between the upload and download costs.

\medskip

\appendix
\section{Appendix}
\subsection{Proof of Corollary \ref{cor:epsilon}} \label{app:corepsilon}
Let us focus on \eqref{eq: epsilon1}, i.e.,  inter-user privacy from the $1^{st}$ user's perspective. Similar reasoning will apply to \eqref{eq: epsilon2}.

When $T_{1} = T_{2} = 1$, $N = L + 2$, we neglect the $t$ on superscripts or subscripts of all symbols. With this simplified notation, the private randomness of each of the two users can be expressed as
\begin{align*}
    \mathcal{Z}_{1} = \{\mathbf{Z}_{1}^{(l)} \mid l \in [1:L]\},&& \mathcal{Z}_{2} = \{\mathbf{Z}_{2}^{(l)} \mid l \in [1:L]\}.
\end{align*}
The intermediate result computed by the $n^{th}$ server can be written as
\begin{align}
    B_{n}^{(\theta_{1}, \theta_{2})} &= \frac{1}{f_{1} - \alpha_{n}}{Q_{n,1}^{(1,\theta_{1})}}^{\prime} \mathbf{W}^{(1)} Q_{n,1}^{(2,\theta_{2})} + \cdots + \frac{1}{f_{L} - \alpha_{n}}{Q_{n,L}^{(1,\theta_{1})}}^{\prime} \mathbf{W}^{(L)} Q_{n,L}^{(2,\theta_{2})}\\
    &= \frac{1}{f_{1} - \alpha_{n}}\mathbf{e}_{K_1}(\theta_{1})^{\prime} \mathbf{W}^{(1)} \mathbf{e}_{K_2}(\theta_{2}) + \cdots + \frac{1}{f_{L} - \alpha_{n}}\mathbf{e}_{K_1}(\theta_{1})^{\prime} \mathbf{W}^{(L)} \mathbf{e}_{K_2}(\theta_{2})\\
    &+ \underbrace{\sum_{l \in [1:L]}\Big({\mathbf{Z}_{1}^{(l)}}^{\prime} \mathbf{W}^{(l)} \mathbf{e}_{K_2}(\theta_{2}) + \mathbf{e}_{K_1}(\theta_{1})^{\prime} \mathbf{W}^{(l)} \mathbf{Z}_{2}^{(l)} + f_{l} {\mathbf{Z}_{1}^{(l)}}^{\prime} \mathbf{W}^{(l)} \mathbf{Z}_{2}^{(l)}\Big)}_{I_{0}}\\
    &+ \alpha_{n} \underbrace{\Big(-\sum_{l \in [1:L]}{\mathbf{Z}_{1}^{(l)}}^{\prime} \mathbf{W}^{(l)} \mathbf{Z}_{2}^{(l)}\Big)}_{I_{1}}.
\end{align}
Note here that even though the expressions for $I_{0}, I_{1}$ are fairly involved, they are just 2 random variables in $\mathbb{F}_{q}$. Meanwhile, $B_{[1:N]}^{(\theta_{1}, \theta_{2})}$ is an invertible function of $\mathbf{W}(\theta_{1}, \theta_{2}), I_{0}, I_{1}$. 

Let us define three sets that contain all the components of $I_{0}, I_{1}$ except $\mathbf{e}_{K_1}(\theta_{1})^{\prime} \mathbf{W}^{(1)} \mathbf{Z}_{2}^{(1)}$ and ${\mathbf{Z}_{1}^{(1)}}^{\prime} \mathbf{W}^{(1)} \mathbf{Z}_{2}^{(1)}$. Specifically,
\begin{align}
    \mathcal{I}_{1} &= \{{\mathbf{Z}_{1}^{(l)}}^{\prime} \mathbf{W}^{(l)} \mathbf{e}_{K_2}(\theta_{2}) \mid l \in [1 : L]\},\\
    \mathcal{I}_{2} &= \{\mathbf{e}_{K_1}(\theta_{1})^{\prime} \mathbf{W}^{(l)} \mathbf{Z}_{2}^{(l)} \mid l \in [2 : L]\},\\
    \mathcal{I}_{3} &= \{{\mathbf{Z}_{1}^{(l)}}^{\prime} \mathbf{W}^{(l)} \mathbf{Z}_{2}^{(l)} \mid l \in [2 : L]\}.
\end{align}

So in $q$-ary units, we have
\begin{align}
    &I(\theta_{2}; B_{[1:N]}^{(\theta_{1}, \theta_{2})} \mid \theta_{1}, \mathcal{Z}_{1}, \mathbf{W}(\theta_{1}, \theta_{2}))\\
    &=I(\theta_{2}; \mathbf{W}(\theta_{1}, \theta_{2}), I_{0}, I_{1} \mid \theta_{1}, \mathcal{Z}_{1}, \mathbf{W}(\theta_{1}, \theta_{2}))\\
    &=I(\theta_{2}; I_{0}, I_{1} \mid \theta_{1}, \mathcal{Z}_{1}, \mathbf{W}(\theta_{1}, \theta_{2}))\\
    &=H(I_{0}, I_{1} \mid \theta_{1}, \mathcal{Z}_{1}, \mathbf{W}(\theta_{1}, \theta_{2})) - H(I_{0}, I_{1} \mid \theta_{1}, \mathcal{Z}_{1}, \mathbf{W}(\theta_{1}, \theta_{2}), \theta_{2})\\
    &\leq 2 - H(I_{0}, I_{1} \mid \theta_{1}, \mathcal{Z}_{1}, \mathbf{W}(\theta_{1}, \theta_{2}), \theta_{2}, \mathcal{I}_{[1:3]}) \label{eq: epsilon3}\\
    &=2 - H(\mathbf{e}_{K_1}(\theta_{1})^{\prime} \mathbf{W}^{(1)} \mathbf{Z}_{2}^{(1)}, {\mathbf{Z}_{1}^{(1)}}^{\prime} \mathbf{W}^{(1)} \mathbf{Z}_{2}^{(1)} \mid \theta_{1}, \mathcal{Z}_{1}, \mathbf{W}(\theta_{1}, \theta_{2}), \theta_{2}, \mathcal{I}_{[1:3]}).\label{eq: epsilon4}
\end{align}
\eqref{eq: epsilon3} results from the fact that $I_{0}, I_{1}$ are in $\mathbb{F}_{q}$ and conditioning reduces entropy. \eqref{eq: epsilon4} holds because elements in $\mathcal{I}_{[1:3]}$ can be subtracted from $I_{0}, I_{1}$.

To proceed further we need to define the following new random variables.
\begin{align}
    E_{1} &= \left\{
    \begin{array}{cc}
    1, &\text{if $\mathbf{W}^{(1)}$ has full-rank},\\
    0, &\text{otherwise}.
    \end{array}
    \right.\\
    E_{2} &= \left\{
    \begin{array}{cc}
    1, &\text{if $\mathbf{Z}_{1}^{(1)} \neq \mathbf{0}$ and $\mathbf{Z}_{1}^{(1)} \indep \mathbf{e}_{K_1}(\theta_{1})$},\\
    0, &\text{otherwise}.
    \end{array}
    \right.
\end{align}
Recall that $\mathbf{Z}_{1}^{(1)} \indep \mathbf{e}_{K_1}(\theta_{1})$ denotes that the two vectors are linearly independent. We have 
\begin{align}
    &\pr(E_{1} = 1) = \frac{\prod_{i \in [1:K]} (q^{K} - q^{i - 1})}{q^{K^{2}}},\label{eq:E1full}\\
    &\pr(E_{2} = 1) = 1 - \frac{1}{q^{K-1}},\\
    &\pr(E_{1} = 1, E_{2} = 1) = \pr(E_{1} = 1) \cdot \pr(E_{2} = 1)\label{eq:E1E2ind}.
\end{align}
Note that the numerator of \eqref{eq:E1full} is the order of the general linear group of degree $K$ over $\mathbb{F}_{q}$. \eqref{eq:E1E2ind} follows because $E_1$ and $E_2$ are independent.

Consider the second term of \eqref{eq: epsilon4}, we have
\begin{align}
    &H(\mathbf{e}_{K_1}(\theta_{1})^{\prime} \mathbf{W}^{(1)} \mathbf{Z}_{2}^{(1)}, {\mathbf{Z}_{1}^{(1)}}^{\prime} \mathbf{W}^{(1)} \mathbf{Z}_{2}^{(1)} \mid \theta_{1}, \mathcal{Z}_{1}, \mathbf{W}(\theta_{1}, \theta_{2}), \theta_{2}, \mathcal{I}_{[1:3]})\\
    \geq&H(\mathbf{e}_{K_1}(\theta_{1})^{\prime} \mathbf{W}^{(1)} \mathbf{Z}_{2}^{(1)}, {\mathbf{Z}_{1}^{(1)}}^{\prime} \mathbf{W}^{(1)} \mathbf{Z}_{2}^{(1)} \mid \theta_{1}, \mathcal{Z}_{1}, \mathbf{W}(\theta_{1}, \theta_{2}), \theta_{2}, \mathcal{I}_{[1:3]}, E_{1}, E_{2})\\
    \geq&\pr(E_{1} = 1, E_{2} = 1)\notag\\
    &H(\mathbf{e}_{K_1}(\theta_{1})^{\prime} \mathbf{W}^{(1)} \mathbf{Z}_{2}^{(1)}, {\mathbf{Z}_{1}^{(1)}}^{\prime} \mathbf{W}^{(1)} \mathbf{Z}_{2}^{(1)} \mid \theta_{1}, \mathcal{Z}_{1}, \mathbf{W}(\theta_{1}, \theta_{2}), \theta_{2}, \mathcal{I}_{[1:3]}, E_{1} = 1, E_{2} = 1).\label{eq: epsilon5}
\end{align}

Let $\mathbf{R}_1, \mathbf{R}_2$ be two row vectors and 
\begin{align}
    \mathbf{R}_{1} = \mathbf{e}_{K_1}(\theta_{1})^{\prime} \mathbf{W}^{(1)}, \mathbf{R}_{2} = {\mathbf{Z}_{1}^{(1)}}^{\prime} \mathbf{W}^{(1)}.
\end{align}
$E_{1} = 1$ implies that $\mathbf{W}^{(1)}$ has full-rank. $E_{2} = 1$ means that $\mathbf{Z}_{1}^{(1)}$ and $\mathbf{e}_{K_1}(\theta_{1})$ are linearly independent. So $\mathbf{R}_{1}, \mathbf{R}_{2}$ are linearly independent. Let $(i, j) \in [1:K]\times[1:K], i \neq j$ be the smallest pair such that
\begin{align}
  \mathbf{M} = \left[
    \begin{array}{c c} 
    \mathbf{R}_{1}(i) & \mathbf{R}_{1}(j)\\
    \mathbf{R}_{2}(i) & \mathbf{R}_{2}(j)
    \end{array}
    \right],
    \det(\mathbf{M}) \neq 0.
\end{align}
Such $(i,j)$ must exist due to the linear independence of $\mathbf{R}_{1}$ and $\mathbf{R}_{2}$.

Let $\overline{\mathcal{Z}_{2}} = \{\mathbf{Z}_{2}^{(1)}(k) \mid k \in [1:K] \setminus \{i, j\}\}$ contain all the entries of $\mathbf{Z}_{2}^{(1)}$ except $\mathbf{Z}_{2}^{(1)}(i), \mathbf{Z}_{2}^{(1)}(j)$, for \eqref{eq: epsilon5}, we have
\begin{align}
    2 \geq &H(\mathbf{e}_{K_1}(\theta_{1})^{\prime} \mathbf{W}^{(1)} \mathbf{Z}_{2}^{(1)}, {\mathbf{Z}_{1}^{(1)}}^{\prime} \mathbf{W}^{(1)} \mathbf{Z}_{2}^{(1)} \mid \theta_{1}, \mathcal{Z}_{1}, \mathbf{W}(\theta_{1}, \theta_{2}), \theta_{2}, \mathcal{I}_{[1:3]}, E_{1} = 1, E_{2} = 1)\\
    =&H(\mathbf{R}_{1} \mathbf{Z}_{2}^{(1)}, \mathbf{R}_{2} \mathbf{Z}_{2}^{(1)} \mid \theta_{1}, \mathcal{Z}_{1}, \mathbf{W}(\theta_{1}, \theta_{2}), \theta_{2}, \mathcal{I}_{[1:3]}, E_{1} = 1, E_{2} = 1)\\
    \geq &H(\mathbf{R}_{1} \mathbf{Z}_{2}^{(1)}, \mathbf{R}_{2} \mathbf{Z}_{2}^{(1)} \mid \theta_{1}, \mathcal{Z}_{1}, \mathbf{W}(\theta_{1}, \theta_{2}), \theta_{2}, \mathcal{I}_{[1:3]}, E_{1} = 1, E_{2} = 1, \mathbf{R}_{1}, \mathbf{R}_{2}, i, j, \overline{\mathcal{Z}_{2}})\\
    =&H\left(\mathbf{M}\left.\left[
        \begin{array}{c}
        \mathbf{Z}_{2}^{(1)}(i)\\
        \mathbf{Z}_{2}^{(1)}(j)
        \end{array} 
        \right] \right| \theta_{1}, \mathcal{Z}_{1}, \mathbf{W}(\theta_{1}, \theta_{2}), \theta_{2}, \mathcal{I}_{[1:3]}, E_{1} = 1, E_{2} = 1, \mathbf{R}_{1}, \mathbf{R}_{2}, i, j, \overline{\mathcal{Z}_{2}}\right)\label{eq: epsilon6}\\
    =&H(\mathbf{Z}_{2}^{(1)}(i), \mathbf{Z}_{2}^{(1)}(j) \mid \theta_{1}, \mathcal{Z}_{1}, \mathbf{W}(\theta_{1}, \theta_{2}), \theta_{2}, \mathcal{I}_{[1:3]}, E_{1} = 1, E_{2} = 1, \mathbf{R}_{1}, \mathbf{R}_{2}, i, j, \overline{\mathcal{Z}_{2}}) = 2\label{eq: epsilon7}
\end{align}
in $q$-ary units. \eqref{eq: epsilon6} holds because we can subtract other components of $\mathbf{R}_{1}\mathbf{Z}_{2}^{(1)}, \mathbf{R}_{2}\mathbf{Z}_{2}^{(1)}$ given the conditioned terms. \eqref{eq: epsilon7} results from the fact that $\mathbf{M}$ is invertible and $\mathbf{Z}_{2}^{(1)}(i)$, $\mathbf{Z}_{2}^{(1)}(j)$ are independent of all conditioned terms.

So for the second term of \eqref{eq: epsilon4} we have 
\begin{align}
    H(\mathbf{e}_{K_1}(\theta_{1})^{\prime} \mathbf{W}^{(1)} \mathbf{Z}_{2}^{(1)}, {\mathbf{Z}_{1}^{(1)}}^{\prime} \mathbf{W}^{(1)} \mathbf{Z}_{2}^{(1)} \mid \theta_{1}, \mathcal{Z}_{1}, \mathbf{W}(\theta_{1}, \theta_{2}), \theta_{2}, \mathcal{I}_{[1:3]}) \geq 2\pr(E_{1} = 1, E_{2} = 1).\label{eq: epsilon8}
\end{align}
Combining \eqref{eq: epsilon8} with \eqref{eq: epsilon4}, we have
\begin{align}
    &I(\theta_{2}; B_{[1:N]}^{(\theta_{1}, \theta_{2})} \mid \theta_{1}, \mathcal{Z}_{1}, \mathbf{W}(\theta_{1}, \theta_{2})) \label{eq: epsilon9}\\
    &\leq 2\Big(1-\pr(E_{1} = 1, E_{2} = 1)\Big)\\
    &= 2 \left( 1 - \left(1 - \frac{1}{q^{K - 1}}\right) \frac{\prod_{k \in [1:K]} (q^{K} - q^{k - 1})}{q^{K^{2}}}\right)\\
    &\leq 2 \left( 1 - \left(1 - \frac{1}{q^{K - 1}}\right) \frac{(q^{K} - q^{K - 1})^K}{q^{K^{2}}}\right)\\
    &= 2 \left( 1 - \left(1 - \frac{1}{q^{K - 1}}\right) \left( 1- \frac{1}{q}\right)^{K}\right). \label{eq: epsilon10}
\end{align}
To ensure that the LHS of \eqref{eq: epsilon9} is bounded above by $\epsilon$ for $q > q_{0}$, we can choose $q_{0}$ to be any value of $q$ that bounds the RHS of \eqref{eq: epsilon10} above by $\epsilon$.
$\hfill\square$

\subsection{Proof of Corollary \ref{cor:XS-TPIR}}\label{app:corXS-TPIR}
 The lower-bound follows already from the proof of achievability of Theorem \ref{thm: rate}. Here we prove the upper bound. Any MB-XS-TPIR scheme with parameters $K_{1}, \cdots, K_{M}, T_{1}, \cdots, T_{M}$ yields a total of $M$ XS-TPIR schemes. For the $m^{th}$ XS-TPIR scheme where $m\in[1:M]$, the user corresponds to the $m^{th}$ user of MB-XS-TPIR. All other users in MB-XS-TPIR generate fixed indices so that the user is retrieving a message in a database with $K_{m}$ messages, i.e., $\mathbf{W}(i_1,\cdots,i_{m-1},\theta_m,i_{m+1},\cdots,i_{M})$ where $\theta_{m} \in [1:K_m]$ while $i_{1},\cdots,i_{m-1},i_{m+1},\cdots,i_M$ are fixed, subject to $T_{m}$-privacy constraint from $N$ servers. The rate of MB-XS-TPIR cannot exceed $\frac{1 - \frac{T_m+X}{N}}{1 - (\frac{T_{m}}{N-X})^{K_m}}$ because this value is the upper bound of the achievable rates of XS-TPIR with $N$ servers, $K_{m}$ messages and $T_{m}$-privacy constraint according to \cite{Jia_Sun_Jafar_XSTPIR}. Since this upper bound holds for all $m\in[1:M]$, the upper bound of \eqref{cor:bound} follows.

\medskip

\bibliographystyle{IEEEtran}
\bibliography{Thesis}

\begin{thebibliography}{10}
\providecommand{\url}[1]{#1}
\csname url@samestyle\endcsname
\providecommand{\newblock}{\relax}
\providecommand{\bibinfo}[2]{#2}
\providecommand{\BIBentrySTDinterwordspacing}{\spaceskip=0pt\relax}
\providecommand{\BIBentryALTinterwordstretchfactor}{4}
\providecommand{\BIBentryALTinterwordspacing}{\spaceskip=\fontdimen2\font plus
\BIBentryALTinterwordstretchfactor\fontdimen3\font minus
  \fontdimen4\font\relax}
\providecommand{\BIBforeignlanguage}[2]{{%
\expandafter\ifx\csname l@#1\endcsname\relax
\typeout{** WARNING: IEEEtran.bst: No hyphenation pattern has been}%
\typeout{** loaded for the language `#1'. Using the pattern for}%
\typeout{** the default language instead.}%
\else
\language=\csname l@#1\endcsname
\fi
#2}}
\providecommand{\BIBdecl}{\relax}
\BIBdecl

\bibitem{Shamir}
A.~Shamir, ``How to share a secret,'' \emph{Communications of the ACM},
  vol.~22, pp. 612--613, 1979.

\bibitem{Gertner_Goldwasser_Malkin}
Y.~Gertner, S.~Goldwasser, and T.~Malkin, ``A random server model for private
  information retrieval,'' in \emph{Randomization and Approximation Techniques
  in Computer Science}.\hskip 1em plus 0.5em minus 0.4em\relax Springer, 1998,
  pp. 200--217.

\bibitem{PIRfirst}
B.~Chor, O.~Goldreich, E.~Kushilevitz, and M.~Sudan, ``Private information
  retrieval,'' in \emph{Proceedings of the 36th Annual Symposium on Foundations
  of Computer Science}, 1995, pp. 41--50.

\bibitem{PIRfirstjournal}
B.~Chor, E.~Kushilevitz, O.~Goldreich, and M.~Sudan, ``Private information
  retrieval,'' \emph{Journal of the ACM (JACM)}, vol.~45, no.~6, pp. 965--981,
  1998.

\bibitem{yao1982protocols}
A.~C. Yao, ``Protocols for secure computations,'' in \emph{Foundations of
  Computer Science, 1982. SFCS'08. 23rd Annual Symposium on}.\hskip 1em plus
  0.5em minus 0.4em\relax IEEE, 1982, pp. 160--164.

\bibitem{Yao_2PC}
A.~C.-C. Yao, ``How to generate and exchange secrets,'' in \emph{27th Annual
  Symposium on Foundations of Computer Science (sfcs 1986)}.\hskip 1em plus
  0.5em minus 0.4em\relax IEEE, 1986, pp. 162--167.

\bibitem{Goldreich_Micali_Wigderson_MPC}
O.~Goldreich, S.~Micali, and A.~Wigderson, ``How to play any mental game, or a
  completeness theorem for protocols with honest majority,'' in \emph{Providing
  Sound Foundations for Cryptography: On the Work of Shafi Goldwasser and
  Silvio Micali}, 2019, pp. 307--328.

\bibitem{Feige_Killian_Naor_PSM}
U.~Feige, J.~Killian, and M.~Naor, ``A minimal model for secure computation,''
  in \emph{Proceedings of the twenty-sixth annual ACM symposium on Theory of
  computing}, 1994, pp. 554--563.

\bibitem{Chan_Ho_Yamamoto}
T.~H. Chan, S.-W. Ho, and H.~Yamamoto, ``Private information retrieval for
  coded storage,'' \emph{Proceedings of IEEE International Symposium on
  Information Theory (ISIT)}, pp. 2842--2846, 2015.

\bibitem{Sun_Jafar_PIR}
H.~Sun and S.~A. Jafar, ``The capacity of private information retrieval,''
  \emph{IEEE Transactions on Information Theory}, vol.~63, no.~7, pp.
  4075--4088, July 2017.

\bibitem{Sun_Jafar_TPIR}
------, ``The capacity of robust private information retrieval with colluding
  databases,'' \emph{IEEE Transactions on Information Theory}, vol.~64, no.~4,
  pp. 2361--2370, April 2018.

\bibitem{Sun_Jafar_SPIR}
------, ``The capacity of symmetric private information retrieval,'' \emph{IEEE
  Transactions on Information Theory}, vol.~65, no.~1, pp. 322--329, 2018.

\bibitem{Samy_Attia_Tandon_Lazos_Weak}
I.~Samy, M.~A. Attia, R.~Tandon, and L.~Lazos, ``Asymmetric leaky private
  information retrieval,'' \emph{arXiv preprint arXiv:2006.03048}, 2020.

\bibitem{Lin_Kumar_Rosnes_Eitan_Weak}
H.-Y. Lin, S.~Kumar, E.~Rosnes, E.~Yaakobi \emph{et~al.}, ``Multi-server
  weakly-private information retrieval,'' \emph{arXiv preprint
  arXiv:2007.10174}, 2020.

\bibitem{Wang_Skoglund_SPIREve}
Q.~Wang and M.~Skoglund, ``Secure private information retrieval from colluding
  databases with eavesdroppers,'' \emph{arXiv preprint arXiv:1710.01190}, 2017.

\bibitem{Wang_Sun_Skoglund}
Q.~Wang, H.~Sun, and M.~Skoglund, ``The capacity of private information
  retrieval with eavesdroppers,'' \emph{IEEE Transactions on Information
  Theory}, vol.~65, no.~5, pp. 3198--3214, 2018.

\bibitem{Wang_Skoglund_PIRSPIRAd}
Q.~Wang and M.~Skoglund, ``On {PIR} and symmetric {PIR} from colluding
  databases with adversaries and eavesdroppers,'' \emph{IEEE Transactions on
  Information Theory}, vol.~65, no.~5, pp. 3183--3197, 2018.

\bibitem{Wang_Sun_Skoglund_BSPIR}
Q.~Wang, H.~Sun, and M.~Skoglund, ``The $\epsilon$-error capacity of symmetric
  {PIR} with {Byzantine} adversaries,'' \emph{arXiv preprint arXiv:1809.03988},
  2018.

\bibitem{Banawan_Ulukus_BPIR}
K.~{Banawan} and S.~{Ulukus}, ``The capacity of private information retrieval
  from {Byzantine} and colluding databases,'' \emph{IEEE Transactions on
  Information Theory}, vol.~65, no.~2, pp. 1206--1219, Feb 2019.

\bibitem{FREIJ_HOLLANTI}
R.~Freij-Hollanti, O.~Gnilke, C.~Hollanti, and D.~Karpuk, ``Private information
  retrieval from coded databases with colluding servers,'' \emph{SIAM Journal
  on Applied Algebra and Geometry}, vol.~1, no.~1, pp. 647--664, 2017.

\bibitem{Tajeddine_Gnilke_Karpuk}
R.~Tajeddine, O.~W. Gnilke, D.~Karpuk, R.~Freij-Hollanti, C.~Hollanti, and
  S.~E. Rouayheb, ``Private information retrieval schemes for coded data with
  arbitrary collusion patterns,'' \emph{arXiv preprint arXiv:1701.07636}, 2017.

\bibitem{Wang_Skoglund_TSPIR}
Q.~Wang and M.~Skoglund, ``Linear symmetric private information retrieval for
  {MDS} coded distributed storage with colluding servers,'' \emph{arXiv
  preprint arXiv:1708.05673}, 2017.

\bibitem{Sun_Jafar_MDSTPIR}
H.~Sun and S.~A. Jafar, ``Private information retrieval from {MDS} coded data
  with colluding servers: Settling a conjecture by {Freij-Hollanti} et al.''
  \emph{IEEE Transactions on Information Theory}, vol.~64, no.~2, pp.
  1000--1022, February 2018.

\bibitem{Tajeddine_Gnilke_Karpuk_Hollanti}
R.~{Tajeddine}, O.~W. {Gnilke}, D.~{Karpuk}, R.~{Freij-Hollanti}, and
  C.~{Hollanti}, ``Private information retrieval from coded storage systems
  with colluding, {Byzantine}, and unresponsive servers,'' \emph{IEEE
  Transactions on Information Theory}, vol.~65, no.~6, pp. 3898--3906, June
  2019.

\bibitem{Wang_Skoglund_MDS}
Q.~Wang and M.~Skoglund, ``Symmetric private information retrieval from {MDS}
  coded distributed storage with non-colluding and colluding servers,''
  \emph{IEEE Transactions on Information Theory}, vol.~65, no.~8, pp.
  5160--5175, 2019.

\bibitem{Jia_Jafar_MDSXSTPIR}
Z.~Jia and S.~A. Jafar, ``{$X$}-secure {$T$}-private information retrieval from
  {MDS} coded storage with {Byzantine} and unresponsive servers,'' \emph{IEEE
  Transactions on Information Theory}, vol.~66, no.~12, pp. 7427--7438, 2020.

\bibitem{Zhou_Tian_Sun_Liu_Min_Size}
R.~Zhou, C.~Tian, H.~Sun, and T.~Liu, ``Capacity-achieving private information
  retrieval codes from {MDS}-coded databases with minimum message size,''
  \emph{IEEE Transactions on Information Theory}, vol.~66, no.~8, pp.
  4904--4916, 2020.

\bibitem{Yang_Shin_Lee}
H.~Yang, W.~Shin, and J.~Lee, ``Private information retrieval for secure
  distributed storage systems,'' \emph{IEEE Transactions on Information
  Forensics and Security}, vol.~13, no.~12, pp. 2953--2964, December 2018.

\bibitem{Jia_Sun_Jafar_XSTPIR}
Z.~{Jia}, H.~{Sun}, and S.~A. {Jafar}, ``Cross subspace alignment and the
  asymptotic capacity of {$X$}-secure {$T$}-private information retrieval,''
  \emph{IEEE Transactions on Information Theory}, vol.~65, no.~9, pp.
  5783--5798, Sep. 2019.

\bibitem{Jia_Jafar_GXSTPIR}
Z.~{Jia} and S.~A. {Jafar}, ``On the asymptotic capacity of {$X$}-secure
  {$T$}-private information retrieval with graph based replicated storage,''
  \emph{IEEE Transactions on Information Theory}, vol.~66, no.~10, pp.
  6280--6296, 2020.

\bibitem{Attia_Kumar_Tandon}
R.~T. Mohamed Adel~Attia, Deepak~Kumar, ``The capacity of private information
  retrieval from uncoded storage constrained databases,'' \emph{arXiv preprint
  arXiv:1805.04104}, 2018.

\bibitem{Wei_Arasli_Banawan_Ulukus_Decentralized}
Y.-P. Wei, B.~Arasli, K.~Banawan, and S.~Ulukus, ``The capacity of private
  information retrieval from decentralized uncoded caching databases,''
  \emph{Information}, vol.~10, no.~12, p. 372, 2019.

\bibitem{Woolsey_Chen_Ji_Storage}
N.~Woolsey, R.-R. Chen, and M.~Ji, ``Private information retrieval from
  heterogeneous uncoded storage constrained databases with reduced
  sub-messages,'' \emph{arXiv preprint arXiv:1904.02131}, 2019.

\bibitem{Guo_Zhou_Tian_Storage}
T.~Guo, R.~Zhou, and C.~Tian, ``New results on the storage-retrieval tradeoff
  in private information retrieval systems,'' \emph{arXiv preprint
  arXiv:2008.00960}, 2020.

\bibitem{Banawab_Arasli_Wei_Ulukus_Heterogeneous}
K.~Banawan, B.~Arasli, Y.-P. Wei, and S.~Ulukus, ``The capacity of private
  information retrieval from heterogeneous uncoded caching databases,''
  \emph{IEEE Transactions on Information Theory}, vol.~66, no.~6, pp.
  3407--3416, 2020.

\bibitem{Tandon_CachePIR}
R.~Tandon, ``The capacity of cache aided private information retrieval,''
  \emph{arXiv preprint arXiv:1706.07035}, 2017.

\bibitem{Wei_Banawan_Ulukus_Side}
Y.-P. Wei, K.~Banawan, and S.~Ulukus, ``The capacity of private information
  retrieval with partially known private side information,'' \emph{IEEE
  Transactions on Information Theory}, vol.~65, no.~12, pp. 8222--8231, 2019.

\bibitem{Wei_Banawan_Ulukus}
Y.~{Wei}, K.~{Banawan}, and S.~{Ulukus}, ``Fundamental limits of cache-aided
  private information retrieval with unknown and uncoded prefetching,''
  \emph{IEEE Transactions on Information Theory}, vol.~65, no.~5, pp.
  3215--3232, May 2019.

\bibitem{Chen_Wang_Jafar_Side}
Z.~Chen, Z.~Wang, and S.~Jafar, ``The capacity of {$T$}-private information
  retrieval with private side information,'' \emph{IEEE Transactions on
  Information Theory}, vol.~66, no.~8, pp. 4761--4773, 2020.

\bibitem{Sun_Jafar_MPIR}
H.~Sun and S.~A. Jafar, ``Multiround private information retrieval: Capacity
  and storage overhead,'' \emph{IEEE Transactions on Information Theory},
  vol.~64, no.~8, pp. 5743--5754, August 2018.

\bibitem{Yao_Liu_Kang_Multiround}
X.~Yao, N.~Liu, and W.~Kang, ``The capacity of multi-round private information
  retrieval from {Byzantine} databases,'' in \emph{2019 IEEE International
  Symposium on Information Theory (ISIT)}.\hskip 1em plus 0.5em minus
  0.4em\relax IEEE, 2019, pp. 2124--2128.

\bibitem{Banawan_Ulukus_MPIR}
K.~Banawan and S.~Ulukus, ``Multi-message private information retrieval:
  Capacity results and near-optimal schemes,'' \emph{IEEE Transactions on
  Information Theory}, vol.~64, no.~10, pp. 6842--6862, 2018.

\bibitem{Shariatpanahi_Siavoshani_Maddah}
S.~P. Shariatpanahi, M.~J. Siavoshani, and M.~A. Maddah-Ali, ``Multi-message
  private information retrieval with private side information,'' \emph{arXiv
  preprint arXiv:1805.11892}, 2018.

\bibitem{Jia_Jafar_SDMM}
Z.~Jia and S.~Jafar, ``On the capacity of secure distributed matrix
  multiplication,'' \emph{arXiv preprint arXiv:1908.06957}, 2019.

\bibitem{Wang_Banawan_Ulukus_PSI}
Z.~Wang, K.~Banawan, and S.~Ulukus, ``Private set intersection: A multi-message
  symmetric private information retrieval perspective,'' \emph{arXiv preprint
  arXiv:1912.13501}, 2019.

\bibitem{Tian_Sun_Chen_Upload}
C.~Tian, H.~Sun, and J.~Chen, ``Capacity-achieving private information
  retrieval codes with optimal message size and upload cost,'' \emph{IEEE
  Transactions on Information Theory}, vol.~65, no.~11, pp. 7613--7627, 2019.

\bibitem{Yao_Liu_Kang_Collusion_Pattern}
X.~Yao, N.~Liu, and W.~Kang, ``The capacity of private information retrieval
  under arbitrary collusion patterns,'' \emph{arXiv preprint arXiv:2001.03843},
  2020.

\bibitem{Li_Gastpar}
S.~Li and M.~Gastpar, ``Single-server multi-message private information
  retrieval with side information,'' \emph{arXiv preprint arXiv:1808.05797},
  2018.

\bibitem{Li_Gastpar_SSMUPIR}
------, ``Single-server multi-user private information retrieval with side
  information,'' in \emph{2018 IEEE International Symposium on Information
  Theory (ISIT)}.\hskip 1em plus 0.5em minus 0.4em\relax IEEE, 2018, pp.
  1954--1958.

\bibitem{Alex_Single_1}
S.~Kadhe, A.~Heidarzadeh, A.~Sprintson, and O.~O. Koyluoglu, ``On an
  equivalence between single-server {PIR} with side information and locally
  recoverable codes,'' \emph{arXiv preprint arXiv:1907.00598}, 2019.

\bibitem{Alex_Single_2}
A.~Heidarzadeh, F.~Kazemi, and A.~Sprintson, ``Capacity of single-server
  single-message private information retrieval with coded side information,''
  2018.

\bibitem{Alex_Single_3}
------, ``Capacity of single-server single-message private information
  retrieval with private coded side information,'' 2019.

\bibitem{Alex_Single_4}
F.~Kazemi, E.~Karimi, A.~Heidarzadeh, and A.~Sprintson, ``Single-server
  single-message online private information retrieval with side information,''
  \emph{arXiv preprint arXiv:1901.07748}, 2019.

\bibitem{Alex_Single_5}
A.~Heidarzadeh, S.~Kadhe, S.~El~Rouayheb, and A.~Sprintson, ``Single-server
  multi-message individually-private information retrieval with side
  information,'' \emph{arXiv preprint arXiv:1901.07509}, 2019.

\bibitem{Latent_PIR}
I.~Samy, M.~A. Attia, R.~Tandon, and L.~Lazos, ``Latent-variable private
  information retrieval,'' \emph{arXiv preprint arXiv:2001.05998}, 2020.

\bibitem{Mirmohseni_Maddah}
M.~Mirmohseni and M.~A. Maddah-Ali, ``Private function retrieval,'' \emph{arXiv
  preprint arXiv:1711.04677}, 2017.

\bibitem{Sun_Jafar_PC}
H.~{Sun} and S.~A. {Jafar}, ``The capacity of private computation,'' \emph{IEEE
  Transactions on Information Theory}, vol.~65, no.~6, pp. 3880--3897, June
  2019.

\bibitem{Mousavi_MaddahAli_Mirmohseni_Innerproduct}
M.~H. Mousavi, M.~A. Maddah-Ali, and M.~Mirmohseni, ``Private inner product
  retrieval for distributed machine learning,'' \emph{arXiv preprint
  arXiv:1902.06319}, 2019.

\bibitem{Obead_Lin_Rosnes_Kliewer_PFC}
S.~A. Obead, H.-Y. Lin, E.~Rosnes, and J.~Kliewer, ``Private function
  computation for noncolluding coded databases,'' \emph{arXiv preprint
  arXiv:2003.10007}, 2020.

\bibitem{Chen_Wang_Jafar_Search}
Z.~Chen, Z.~Wang, and S.~A. Jafar, ``The asymptotic capacity of private
  search,'' \emph{IEEE Transactions on Information Theory}, vol.~66, no.~8, pp.
  4709--4721, 2020.

\bibitem{Jia_Jafar_CDBC}
Z.~Jia and S.~Jafar, ``Cross-subspace alignment codes for coded distributed
  batch computation,'' \emph{arXiv preprint arXiv:1909.13873}, 2019.

\bibitem{Sun_Jafar_LDC}
H.~Sun and S.~A. Jafar, ``On the capacity of locally decodable codes,''
  \emph{IEEE Transactions on Information Theory}, vol.~66, no.~10, pp.
  6566--6579, 2020.

\bibitem{Ishai_Kushilevitz}
Y.~Ishai and E.~Kushilevitz, ``On the hardness of information-theoretic
  multiparty computation,'' in \emph{Advances in Cryptology-EUROCRYPT
  2004}.\hskip 1em plus 0.5em minus 0.4em\relax Springer, 2004, pp. 439--455.

\bibitem{Kakar_Ebadifar_Sezgin_CSA}
J.~Kakar, S.~Ebadifar, and A.~Sezgin, ``{On the Capacity and
  Straggler-Robustness of Distributed Secure Matrix Multiplication},''
  \emph{IEEE Access}, vol.~7, pp. 45\,783--45\,799, 2019.

\bibitem{Chen_Jia_Wang_Jafar_NGCSA}
Z.~Chen, Z.~Jia, Z.~Wang, and S.~A. Jafar, ``{GCSA} codes with noise alignment
  for secure coded multi-party batch matrix multiplication,'' \emph{arXiv
  preprint arXiv:2002.07750}, 2020.

\bibitem{Cadambe_Grover_Tutorial}
V.~Cadambe and P.~Grover, ``Codes for distributed computing: A tutorial,''
  \emph{IEEE ITSOC Newsletter}, vol.~67, no.~4, pp. 3--15, December 2017.

\bibitem{Yu_Lagrange}
Q.~Yu, S.~Li, N.~Raviv, S.~M.~M. Kalan, M.~Soltanolkotabi, and S.~Avestimehr,
  ``{Lagrange Coded Computing: Optimal Design for Resiliency, Security and
  Privacy},'' \emph{ArXiv:1806.00939}, 2018.

\bibitem{Yu_Maddah-Ali_Avestimehr_Polynomial}
Q.~Yu, M.~A. Maddah-Ali, and A.~S. Avestimehr, ``{Polynomial Codes: an Optimal
  Design for High-Dimensional Coded Matrix Multiplication},'' \emph{arXiv
  preprint arXiv:1705.10464}, 2017.

\bibitem{Habil_Double_Sequence}
E.~D. Habil, ``Double sequences and double series,'' \emph{IUG Journal of
  Natural Studies}, vol.~14, no.~1, 2016.

\bibitem{Gasca_Martinez_Muhlbach}
M.~Gasca, J.~Martinez, and G.~M{\"u}hlbach, ``Computation of rational
  interpolants with prescribed poles,'' \emph{Journal of Computational and
  Applied Mathematics}, vol.~26, no.~3, pp. 297--309, 1989.

\bibitem{Lu_Platanioties_Venetsanopoulos_Tensor}
H.~Lu, K.~N. Plataniotis, and A.~Venetsanopoulos, \emph{Multilinear Subspace
  Learning: Dimensionality Reduction of Multidimensional Data}.\hskip 1em plus
  0.5em minus 0.4em\relax CRC press, 2013.

\bibitem{Wang_Skoglund}
Q.~Wang and M.~Skoglund, ``Symmetric private information retrieval for {MDS}
  coded distributed storage,'' \emph{arXiv preprint arXiv:1610.04530}, 2016.

\bibitem{Wang_Skoglund_SPIRAd}
------, ``Secure symmetric private information retrieval from colluding
  databases with adversaries,'' \emph{arXiv preprint arXiv:1707.02152}, 2017.

\end{thebibliography}

\end{document}